\pdfoutput=1

\documentclass[reprint,twocolumn]{article}

\usepackage{graphicx}
\usepackage{amsmath,amssymb,amsfonts}
\usepackage{color}
\usepackage{xr}
\usepackage{float}
\usepackage{bm}% bold math
\usepackage[colorlinks,linkcolor=blue,citecolor=blue]{hyperref}
\usepackage{color}

\usepackage{epsf}

\usepackage{dcolumn}% Align table columns on decimal point

\usepackage[english]{babel}
\usepackage{url}
\usepackage{bbm}
\usepackage{multirow}

\usepackage{chngcntr}

\usepackage{appendix}

\usepackage{color}

\usepackage{morefloats}

\usepackage{authblk}

\graphicspath{ {figures/} }
\usepackage{array}
\usepackage[utf8]{inputenc}

\DeclareMathOperator*{\argmax}{arg\,max}

\begin{document}

% ******************************************************************************************************************************************
% ******************************************************************************************************************************************
% TITLE
% ******************************************************************************************************************************************
% ******************************************************************************************************************************************

\title{\textbf{Multi-scale analysis and clustering of co-expression networks}}

% ******************************************************************************************************************************************
% ******************************************************************************************************************************************
% AUTHOR INFORMATION
% ******************************************************************************************************************************************
% ******************************************************************************************************************************************

\author[1]{Nuno R. Nen\'{e}\thanks{nn276@cam.ac.uk, nunonene@gmail.com}}
\affil[1]{Department of Genetics, University of Cambridge, Cambridge, UK}

\maketitle

% ******************************************************************************************************************************************
% ******************************************************************************************************************************************
% ABSTRACT
% ******************************************************************************************************************************************
% ******************************************************************************************************************************************

\begin{abstract}
The increasing capacity of high-throughput genomic technologies for generating time-course data has stimulated a rich debate on the most appropriate methods to highlight crucial aspects of data structure. In this work, we address the problem of sparse co-expression network representation of several time-course stress responses in {\it Saccharomyces cerevisiae}. We quantify the information preserved from the original datasets under a graph-theoretical framework and evaluate how cross-stress features can be identified. This is performed both from a node and a network community organization point of view. Cluster analysis, here viewed as a problem of network partitioning, is achieved under state-of-the-art algorithms relying on the properties of stochastic processes on the constructed graphs. Relative performance with respect to a metric-free Bayesian clustering analysis is evaluated and possible extensions are discussed. We further cluster the stress-induced co-expression networks generated independently by using their community organization at multiple scales. This type of protocol allows for integration of multiple datasets that may not be immediately comparable, either due to diverse experimental variations or because they represent different types of information about the same genes.
\end{abstract}

% ******************************************************************************************************************************************
% ******************************************************************************************************************************************
% INTRODUCTION
% ******************************************************************************************************************************************
% ******************************************************************************************************************************************

\addtocontents{toc}{\protect\setcounter{tocdepth}{0}}

\addtocontents{lof}{\protect\setcounter{tocdepth}{0}}
\addtocontents{lot}{\protect\setcounter{tocdepth}{0}}

\section{Introduction}

\begin{figure*}[!tb]
\centering
\includegraphics[width=1\textwidth]{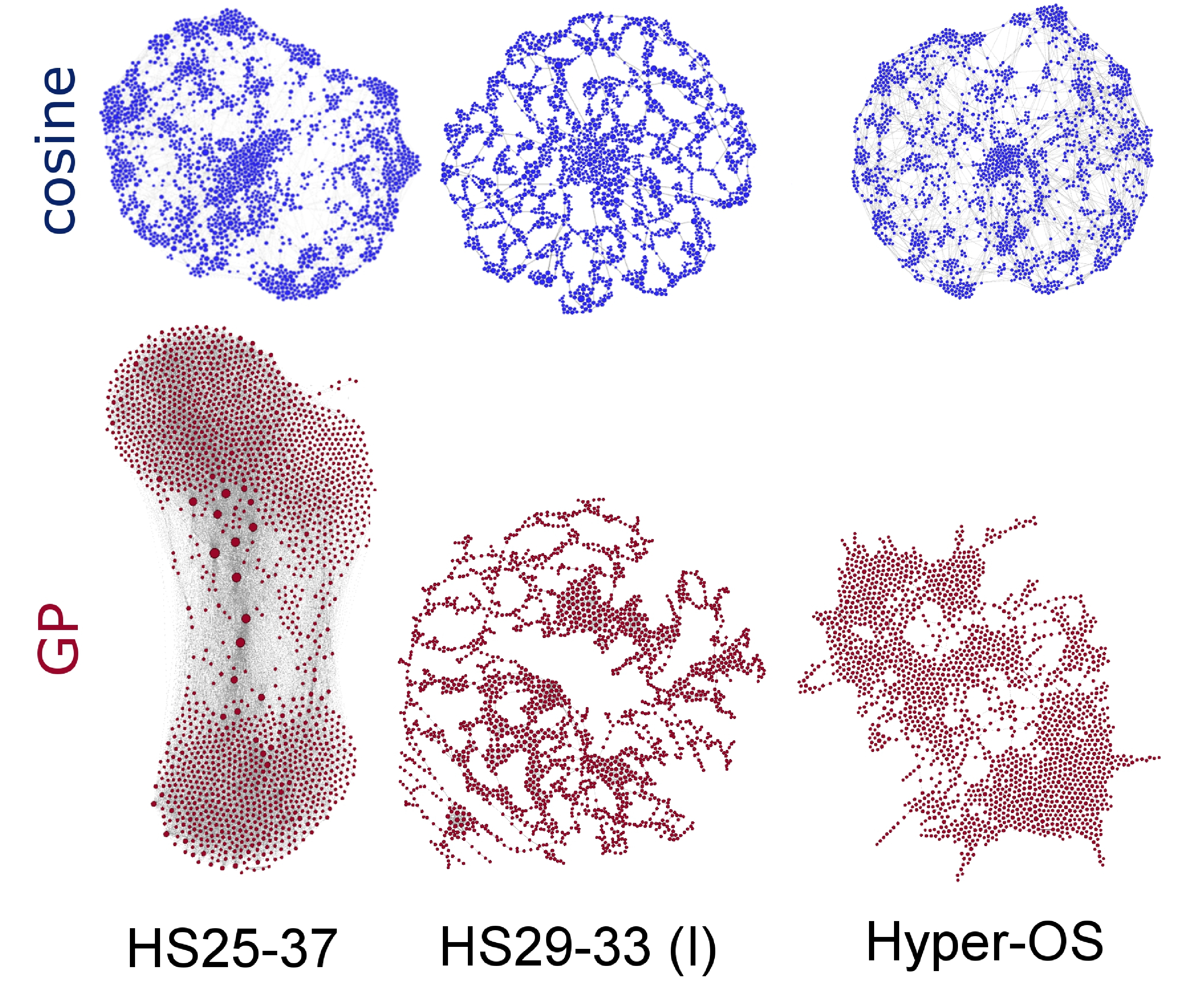}
\caption{Co-expression networks for a selection of stresses for two dissimilarity functions. $HS25-37$: heat-shock from a temperature of 25$^{\circ}$C to 37$^{\circ}C$. $HS29-33 (I)$: mild heat-shock from 29$^{\circ}$C to 33$^{\circ}$C. $Hyper-OS$: hyper-osmotic stress. These networks were plotted via a standard spatial embedding force-based algorithm available in Gephi. See Methods for details.}
\label{fig:Fig_RMSTs}
\end{figure*}

With the advent of technologies allowing the collection of time-series in cell biology, the rich structure of the paths that cells take in expression space became amenable to processing. Several methodologies have been crucial to carefully organize the wealth of information generated by experiments \cite{Bar-Joseph2012}, including network-based approaches which constitute an excellent and flexible option for systems-level understanding \cite{Bar-Joseph2012,Horvath2008,Ruan2010}. The use of graphs for expression analysis is inherently an attractive proposition for reasons related to sparsity, which simplifies the cumbersome analysis of large datasets, in addition to being mathematically convenient (see examples in Fig.~\ref{fig:Fig_RMSTs}). The properties of co-expression networks might reveal a myriad of aspects pertaining to the impact of genes \cite{Horvath2008}, as well as group features, such as structural clusters or communities, highlighting similar expression profiles. The structure of co-expression networks can also be compared with other types of networks, e.g. protein-protein interaction, genetic interaction or gene regulatory, by using simple and scalable graph-theoretical methods such as the one explored here. Additionally, the use of graphs may also help to shed light on the evolutionary differences or commonalities between cellular responses across species \cite{Berg2006,Tirosh2011,Kolar2012}.

% ------------------
% ------------------

% ------------------
% ------------------

Clustering methodologies have been an invaluable tool in unravelling the structure of time-course data \cite{Bar-Joseph2012,Bar-Joseph2003}. Either by using standard approaches relying on cosine dissimilarities and hierarchical clustering \cite{Gasch2000,Gasch2002} or by adopting a Bayesian framework without choosing {\it a priori} a particular distance function \cite{Heard2005,Windram2012,Cooke2011,Kirk2012}, most of the techniques rely on the assumption that the necessary standardization methods that precede the clustering analysis are sufficient to eliminate possible inconsistencies between datasets. Here, we will resort to a graph-theoretical based protocol that is able to avoid the pitfalls of experimental variation. This protocol is based on ideas related to community detection on graphs \cite{Fortunato2010,Lancichinetti2010}, which makes it possible to address the characterization of expression dynamics with methods that take into account multi-level organization information. Overall, the method explores aspects of probability flow and containment across a network \cite{Delvenne2010,Delvenne2013}. Although several other algorithms that rely on similar principles are available, the one we test here is very fast. This makes the analysis of large datasets such as those used in the present work a feasible task. In conjunction with this approach, we also resort to methodologies stemming from the field of non-linear dimensionality reduction. The additional treatment has the objective of successfully identifying relationships between genes that better represent the underlying high-dimensional geometry of the data \cite{Beguerisse-Diaz2014} and, naturally, create sparse and manageable datasets. We test the protocol with two dissimilarity functions capturing different features of the expression dynamics. Each leads to differentiated network representations of the same patterns. This allow us to verify if additional information improves performance of the protocol under different metrics.

The data used in this work is a compendium of stress-induced expression microarrays for {\it S.cerevisiae}, originally published by Gasch and co-workers \cite{Gasch2000}. We chose this dataset due to the variety of stresses applied and the fact that it is still one of the biggest datasets for time-course data, in addition to being a benchmark for testing inference of networks in general \cite{Ruan2010,Lee2002}. The work published in \cite{Gasch2000} includes both single and serial stress responses, and one experiment with a combinatorial-like stress. The subject of combinatorial stress in yeasts has been further explored in more recent work (see for example \cite{Kaloriti2012}). 

The results of Gasch and co-workers \cite{Gasch2000} were a seminal contribution to the analysis of whole-genome expression and identified crucial features of stress response in yeast. Specifically, the presence of a Common or Environmental Stress Response (ESR), also a hallmark in several other species \cite{Tirosh2011,Gasch2007,Brown2014}, was extracted by hierarchical clustering analysis and postulated to be fundamental in equipping {\it S.cerevisiae} to combat serial and combinatorial stress. More recently, the ESR feature has been further analysed and studies have concluded that there is more structure to the signals measured than previously expected \cite{Berry2008,Berry2011}.

The paper is organized as follows: we initially quantify how much information the network construction algorithm used in our work retains from the whole set of selected genes, under different co-expression metrics (section~\ref{sec:co-expression network properties}); this is achieved by evaluating both structural and diffusion properties of each network; following this analysis the constructed networks serve as the basis for the study of diffusion patterns across its structures, at different time-scales, with the intent of identifying clusters of similar time-dependent behaviour (section~\ref{sec:Flow-basedClusteringAnalysis}); finally, all the partition solutions identified for each stress are clustered once again, with respect to other stresses, by taking into account the geometry in partition space (section~\ref{sec:Network comparison}). This sequential analysis allowed us to compare how genes are clustered across stresses and ultimately to identify a joint clustering solution. We also compare the multi-resolution organization of all the co-expression networks with those of protein-protein and genetic interaction networks (section~\ref{sec:BIOGRID}) extracted from $BioGRID$ ($http://thebiogrid.org$). This overall analysis, under the graph-theoretical paradigm, leads to efficient multi-dataset integration. The work-flow underlying the protocol explored here is presented in Fig.~\ref{fig:Flow_diagram}.

% ******************************************************************************************************************************************
% ******************************************************************************************************************************************
% RESULTS
% ******************************************************************************************************************************************
% ******************************************************************************************************************************************

\section{Results}

\subsection{Identification of stress-induced co-expression networks} \label{sec:co-expression nets}

A gene co-expression network is an undirected graph where nodes represent genes and edges proximity between expression profiles \cite{Xulvi-Brunet2010,Song2012}. It does not include or attempt to explicitly represent regulatory or physical interactions, but simply highlights common features of the data pertaining to the nodes it connects. Traditionally, co-expression networks have been used to determine co-expression modules or sets of genes that exhibit similar patterns under the considered experimental conditions. Usually, the full co-expression matrix under a specific dissimilarity function is used or a significance threshold is imposed, either under a specific density kernel or simply ad hoc. The latter leads to removal of links that under the chosen model are not significant and increases speed and performance in the clustering step \cite{Song2012}. Here, we resort to a method for sparsifying co-expression matrices that avoids imposing statistical thresholds to attain edge significance. This method has been tested widely across multiple datasets and provides a robust starting point \cite{Beguerisse-Diaz2014,BorislavVangelov2014}.

Throughout this work a co-expression network will be represented by $G=\lbrace V,E,W \rbrace$, where $V$ represents the set of vertices or nodes/genes, $E$ represents the set of edges or links between genes in matrix form, in our case determined by a particular algorithm \cite{Beguerisse-Diaz2014}, and $W$ represents the set of weights associated with each of the edges in $E$, again in matrix form. The network $G$ derived for each stress will represent the overall intra-stress relationships in the expression time-series matrix, $\mathbf{g(t)}$, for all genes selected for the study, measured by microarray data collected in previous publications \cite{Gasch2000}. The edges in $G$, and respective weights, are derived from a distance matrix, $D$, for specific pair-wise distance or dissimilarity functions, in our case based on the cosine between the expression vectors, referred to as $Cosine$ dissimilarity, or on a likelihood ratio found by Gaussian process regression, referred to as $GP$ dissimilarity (see section~\ref{sec:Dist_fun}).

% ----------------------
% ----------------------
\begin{figure*}[!tb]
\includegraphics[width=1\textwidth]{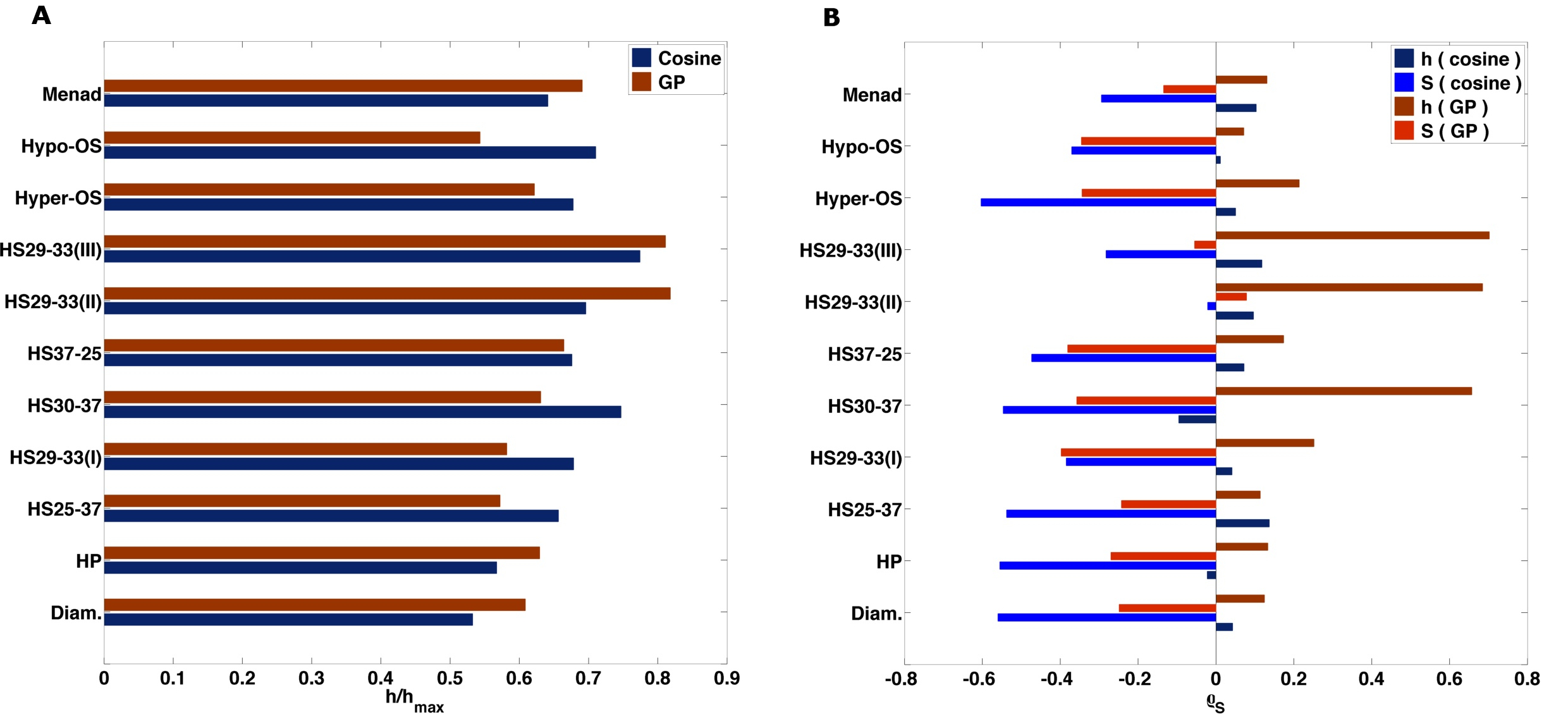}
\caption{Entropy rate for each stress-induced co-expression network. (A) Entropy rate normalized by the optimal entropy rate corresponding to each weighted structure. (B) Rank correlation between entropy rates per node, not-normalized ($-\sum_{j}\pi_{i}p_{ij}log(p_{ij})$ in Eq.~\ref{eq:Eq_EntropyRate}), associated with full weights matrices and corresponding weighted RMSTs.}\label{fig:Fig_EntropyRate}
\end{figure*}
% ----------------------
% ----------------------

In order to represent as faithfully as possible both local and global aspects of the expression dynamics characterizing each gene, we applied a computational technique based on minimum spanning trees \cite{Jackson2010,Jackson2010a}, here referred to as Relaxed Minimum Spanning Tree (RMST) algorithm (see \cite{Beguerisse-Diaz2013,Beguerisse-Diaz2014,BorislavVangelov2014} and the Methods section for details). This approach stems from the field of non-linear dimensionality reduction and manifold reconstruction, where some of the methods rely on the construction of networks that connect only nearest neighbours \cite{Tenenbaum2000,Roweis2000,Carreira2005,Coifman2005}. It allows us to enhance continuity features in the expression profiles collected and secure that the clustering routines proposed here work efficiently. Although minimum spanning trees have been used to sparsify co-expression matrices in the past \cite{Xu2002,Yu2015}, the RMST algorithm provides additional information pertaining to the global geometry of the data. This is the first application of the algorithm in question to time-course data and it contributes to the core  group of graph-based approaches that have performed very well when applied to time-series analysis in biology (see, for example, several applications in \cite{Sharan2003,Liu2014}). Unlike some of the traditional methods for manifold identification, the RMST algorithm  avoids the problem of under-sampling that hinders other methodologies (see for example \cite{Carreira2005} where this is thoroughly addressed). 

In the following section we will highlight several structural properties of the generated co-expression graphs, for both distance functions mentioned above. We also evaluate the information that is preserved from the original pair-wise distance matrix between genes. We focus our analysis on a restricted set of genes selected by appropriate filtering methods that take into account the order of events as a contributing factor. The final set of genes allowed for a multi-scale analysis of stress-specific responses, but also for common features such as the ESR response to be present (see section~\ref{sec:Flowdiagram}, where details on the analysis workflow are graphically depicted).

%--------------------------------------------------------------------------------------------------------------------------------------------
% ------------------------------------------------------------------------------------------------------------------------------------------

\subsubsection{Sparsification and preservation of information content under different dissimilarity functions.} \label{sec:co-expression network properties}

Sparsification of the distance matrices used in the creation of co-expression networks eliminates redundant information. As stated above, the RMST algorithm aims at preserving only links that definitely represent close relationships between neighbour genes. If each stress has a very distinctive associated response, captured by either of the distance functions used in our work, we expect this to be clear in diverse structural properties, either pertaining to graph density, centrality or community organization. Some of the networks can be visualized in Fig.~\ref{fig:Fig_RMSTs}. At first sight we can observe that they are very different in their spatial embedding. In fact, a simple hierarchical clustering of all of the networks based on a Jaccard distance, which measures the number of shared edges, further reveals this (see Supplementary Fig.~\ref{fig:Fig_Hclust_Jaccard}); the minimum value achieved  between any of the stresses is around 0.98, which further confirms the disparity between stress-induced networks.  Yet, the effect of the dissimilarity functions tested in this work can already be seen in the hierarchy of groups along the dendrograms. For example, different classes of stresses tend to be clustered together.

Regarding connectivity properties, we can observe that the graph density (equated with $\frac{\sum_{i}\sum_{j\neq i} e_{ij}}{n(n-1)}$, where $e_{ij}$ represents the Boolean adjacency matrix ($E$) entries and $n$ its dimension) and degree distributions are considerably different across stresses and distance functions (see Supplementary Figs.~\ref{fig:Fig_RMSTsDensity} and \ref{fig:Fig_RMSTsDegree}). For the cosine dissimilarity the largest densities are associated with the stresses menadione ($Menad.$), a superoxide-generating drug, diamide ($Diam.$), a sulfhydryl-oxidizing agent, hydrogen peroxide ($HP$) and heat-shock ($HS25-37$), in decreasing order. The difference observed in graph density points to there being characteristic features of each stress resulting in a geometry of expression space not shared by all. Further information on typical connectivity properties are explored in Supplementary Information (section~\ref{sec:AllNetProp}), where stress-specific as well as cross-stress conserved aspects are evaluated in detail. Below, we evaluate diffusion properties on each graph that summarize the information preserved from the original distance matrices, from a global point of view. These measures are consistent with the flow-based community detection framework reported in a further section.

The dynamical properties of a diffusion process over a co-expression graph can be characterized by evaluating the entropy rate (see Eq.~\ref{eq:Eq_EntropyRate}) of the respective ergodic Markov chain. Its value translates how the entropy of the process increases with diffusion time and represents the minimal amount of information necessary to describe the diffusion process on the weighted structure under study. This function has the potential to distinguish between cellular states \cite{Banerji2013}. As in \cite{Gomez-Gardenes2008}, let us first consider a discrete time Markov chain $p_{t+1}=D_{d}^{-1}A p_{t}$, where $D_{d}$ represents the diagonal weighted-degree matrix and $A$ the weighted adjacency matrix:

% ----------------------
% ----------------------
\begin{equation}  
h=-\sum_{i,j}\pi_{i}p_{ij}log(p_{ij})\label{eq:Eq_EntropyRate}
\end{equation}
% ----------------------
% ----------------------

\noindent In Eq.~\ref{eq:Eq_EntropyRate}, $p_{ij}$ represents the entries in the probability transition matrix associated with the discrete-time Markov chain defined on the graph. In this instance, $p_{ij}$ is defined through the weighted edges of the corresponding co-expression network: $p_{ij}=\frac{w_{ij}}{\sum_{ij}w_{ij}}$. In addition, $\pi_{i}=[\mathbf{d}^{T}/2m]_{i}$ is the $i$th component of the stationary distribution associated with the discrete Markov chain previously defined, with $\mathbf{d}$ standing for the vector of weighted-degrees, and $m$ the total weight of the network. 

Since the entropy rate has been proven to depend on the connectivity of the underlying graph \cite{Gomez-Gardenes2008}, in order to compare each of the stress-induced co-expression networks in terms of the diffusion properties we normalized the entropy rate obtained by that associated with a maximal dispersion \cite{Gomez-Gardenes2008,Banerji2013}. The normalization factor to be employed here for each $h$ is the entropy rate associated with a random walk where the stochastic matrix is taken to be $p_{ij}^{ec}=\frac{I(w_{ij}>0)v_{j}}{\lambda v_{i}}$, where $I(w_{ij})$ is the indicator matrix, and $\lambda$ and $\mathbf{v}$ are the largest eigenvalue of $I(w_{ij})$ and the respective eigenvector \cite{Gomez-Gardenes2008,Delvenne2011,Ochab2012}. The stationary probability distribution for a random walk generated by the  $[p_{ij}^{ec}]$ transition matrix is given by $\pi^{ec}=\frac{v_{i}^{2}}{\sum v_{i}^{2}}$. 

Observing the results plotted in Fig.~\ref{fig:Fig_EntropyRate} (A) and Supplementary Fig.~\ref{fig:Heterogeneity_allstresses_CV}, we can notice that the trend for the normalized entropy rate across stresses follows roughly what was expected, i.e. the larger the heterogeneity among weights the more constrained the diffusion across the network and, naturally, the smaller the normalized value for the entropy rate (for cosine dissimilarity the Spearman correlation coefficient $\rho_{S}=-0.83$ and $p-value<<0.05$, for $GP$ dissimilarity $\rho_{S}=-0.51$ and $p-value>0.05$). Overall, the standard cosine dissimilarity generates  entropy rate values that are higher than those obtained for the GP distance function in 6 of the stresses. This indicates, although not strikingly, that from a global perspective, the cosine dissimilarity is less discriminative in the features it represents and is less effective in identifying local structure consistent with the heuristic represented by Eq.~\ref{eq:RMSTalgo} \footnote{ The same tendency is not observed for full weighted matrices (see Supplementary Fig.~\ref{fig:Fig_EntropyRate_Suppl} (B))}. In fact, the results shown for global heterogeneity (see Supplementary Fig.~\ref{fig:Heterogeneity_allstresses_CV}) clearly indicate that, despite the weights retained after sparsification having a significantly lower coefficient of variation, the information encoded in local structure allows the $GP$ to be more specific in its representation of stress responses with less edges (Supplementary Fig.~\ref{fig:Fig_RMSTsDensity}). As will be reported further ahead, this distance function outperforms the $Cosine$ dissimilarity. We should also remark that the largest normalized entropy rates are associated with the combinatorial stresses, $HS29-33 (II)$ and $HS29-33 (III)$. These two contrasting stresses are exactly those that generate networks with considerable density differences within and between distance functions.

In order to further quantify the amount of information retained from each application of the RMST algorithm to each co-expression matrix, we calculated the Spearman correlation between the entropy rate terms ($-\sum_{j}\pi_{i}p_{ij}log(p_{ij})$ in Eq.~\ref{eq:Eq_EntropyRate}) corresponding to genes in the full matrix, which should highlight aspects related to local entropy or heterogeneity of weights, and the entropy rate terms calculated for the final co-expression networks. The latter incorporates the final degree of each node, via the stationary distribution $\mathbf{\pi}$ of the chosen Markov chain, and the heterogeneity of locals weights represented by the local node Shannon entropy. In Fig.~\ref{fig:Fig_EntropyRate} (A), we can observe that for all stresses the correlation coefficients between the ranks obtained for the co-expression network and the full matrix, when a $GP$ distance function is used, are always positive. The results pertaining to the cosine distance function are not as uniform: the stresses $HP$ and $HS30-37$ induce anti-correlated ranks between RMST and full matrix. One interesting aspect about the Spearman correlation calculated for different distance functions is the significance associated with each value (see p-values in Supplementary Fig.~\ref{fig:Fig_EntropyRate_Suppl} (D)). The Gaussian process inspired distance function attains always much larger $-log(p-values)$, and thus higher significance.

% ----------------------
% ----------------------

\begin{figure*}[!tb]
\begin{center}
\includegraphics[width=1\textwidth]{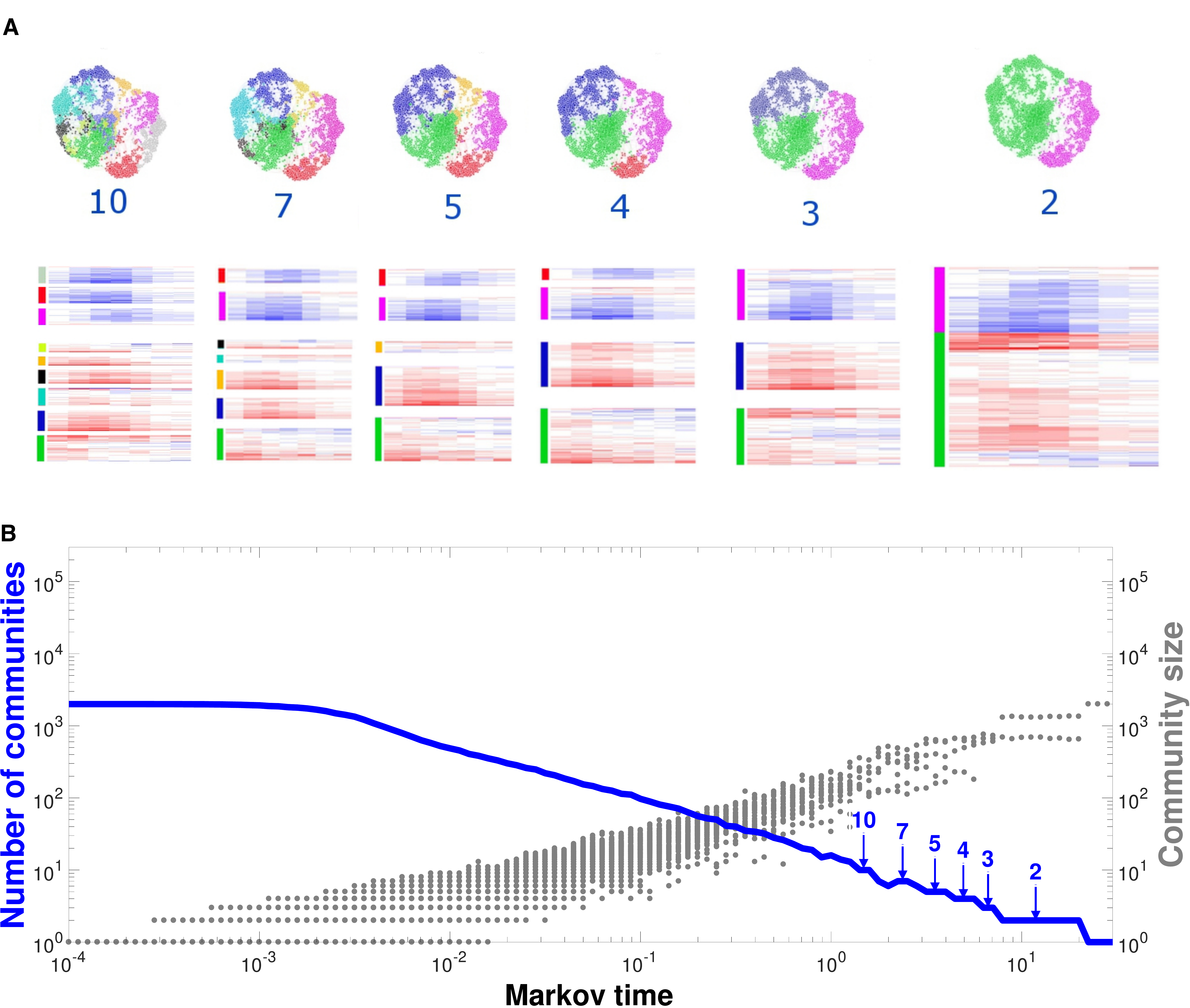}
\caption{Flow-based communities for the co-expression network under heat-shock from a temperature of 25$^{\circ}$C to 37$^{\circ}C$ (HS25-37). (A) Spatial layout (Gephi) with the respective communities identified by color. Expression matrices associated with each cluster represented on the networks are also provided (red induced, blue repressed). Columns of heat maps correspond to expression patterns at 5, 10, 15, 20, 40, 60 and 80 mins. (B) Number and size of communities with Markov time. Dissimilarity function: Cosine.}
\label{fig:Fig_StabilitySolutionHS25-37}
\end{center}
\end{figure*}
% ----------------------
% ----------------------

We also provide another rank measure associated with each node which is essentially a normalized non-equilibrium entropy score, $S_{i}=-\frac{\sum_{j} p_{ij}log(p_{ij})}{log k_{i}}$ \cite{West2012}. In contrast to the entropy rate, it does not dependent on the stationary distribution ${\bf \pi}$ of the discrete time Markov chain defined above. The only aspect that is salient, comparing with results previously highlighted for the entropy rate, is the fact that for both  distance functions all correlation coefficients are negative and also very significant. This shows that the application of the RMST algorithm induces a drastic reordering of the ranks associated with the non-equilibrium entropy calculated for the respective full matrix (Supplementary Fig.~\ref{fig:Fig_EntropyRate_Suppl} (B)). This calculation indicates that the local heterogeneity is substantially changed during the sparsification procedure.

Also of special interest for cross-stress clustering analysis, is the fact that the genes belonging to the common stress response highlighted before have higher cross-stress entropy rate, and non-equilibrium rank correlations, than the rest of the network (see Supplementary Fig.~\ref{fig:Fig_SpearmanCorr_allexp_boxplot}). This further emphasizes the similarity of their expression profiles across stresses and encourages a view of each co-expression network from a community organization point of view (further explored in the following section).

% ------------------------------------------------------------------------------------------------------------------------------------------
% ------------------------------------------------------------------------------------------------------------------------------------------
\subsection{Flow-based clustering analysis: multi-resolution characterization of expression patterns} \label{sec:Flow-basedClusteringAnalysis}

The representation capacity of the co-expression networks for each of the distance functions is, ultimately, as was mentioned above, a tool for clustering analysis. Although structures differ substantially across stresses and distance functions, the diffusion properties of dynamical processes on each structure may reveal similar partitions and, consequently, allow us to reduce the information pertaining to the response under each stress by organizing it into clusters of genes with similar responses. In fact, as was thoroughly described in \cite{Onnela2012,Lancichinetti2010}, a mesoscopic level of description might reveal additional shared features of the ensemble of networks. Here, instead of relying exclusively on state variables such as entropy \cite{Onnela2012}, we compare clustering solutions revealed by a diffusion-based quality function defined on the graph (see Eq.~\ref{eq:Eq_stability}). 

The multi-resolution view has been proven to reveal important features of networks not captured by other methodologies. Despite the appeal of this approach, most alternatives to the method proposed here compute a candidate for the best partition that suffers from an over-partitioning problem \cite{Schaub2012,Delvenne2013}. This is also a characteristic  of clustering methodologies tailored for gene expression time-series analysis \cite{Heard2005,Cooke2011}. Markov stability analysis allows, on the other hand, for evaluating the way communities or clusters group to generate coarser partitions highlighting higher levels of biological function \cite{Delmotte2011}. 

As can be visualized in Fig.~\ref{fig:Fig_StabilitySolutionHS25-37} (B), the partitions found by Markov stability analysis become coarser with Markov time $t$. Each community or cluster becomes larger and larger as the dynamical process on the network spans larger and larger portions. In the same picture, we have plotted the corresponding co-expression networks for the partitions indicated in blue, i.e. composed of 10, 7, 5, 4, 3 and 2 clusters. The identified clusters are plotted in different colours in each respective partition solution. These partitions were selected according to robustness criteria relying on the calculation of the variation of information between 100 possible solutions of a greedy search algorithm, the Louvain algorithm \cite{Blondel2008}. As it is evident, the patterns of expression for each of the clusters in each solution are successfully grouped from partition 10 to partition 2, the latter corresponding to an overall rough division between differentially induced (red) and repressed (blue) genes. Another confirmation of the successful clustering analysis performed here is the spatial embedding plotted in the same figure; each of the network nodes that are collocated have similar colours and the gradual coarsening of the partitions follows from agglomeration of adjacent clusters. We must emphasize that the application of Markov stability analysis, through the Louvain algorithm, does not impose a strict hierarchical grouping of clusters: for each Markov time the best partition is found among 100 candidates without taking into account partitions at previous times. Therefore, the hierarchical organization of the sequence of partitions follows naturally from the diffusion properties \cite{Delvenne2013,Lambiotte2015}, something not available in other techniques for time-series analysis.

% ----------------------
% ----------------------

\begin{figure*}[!tb]
\begin{center}
\includegraphics[width=1\textwidth]{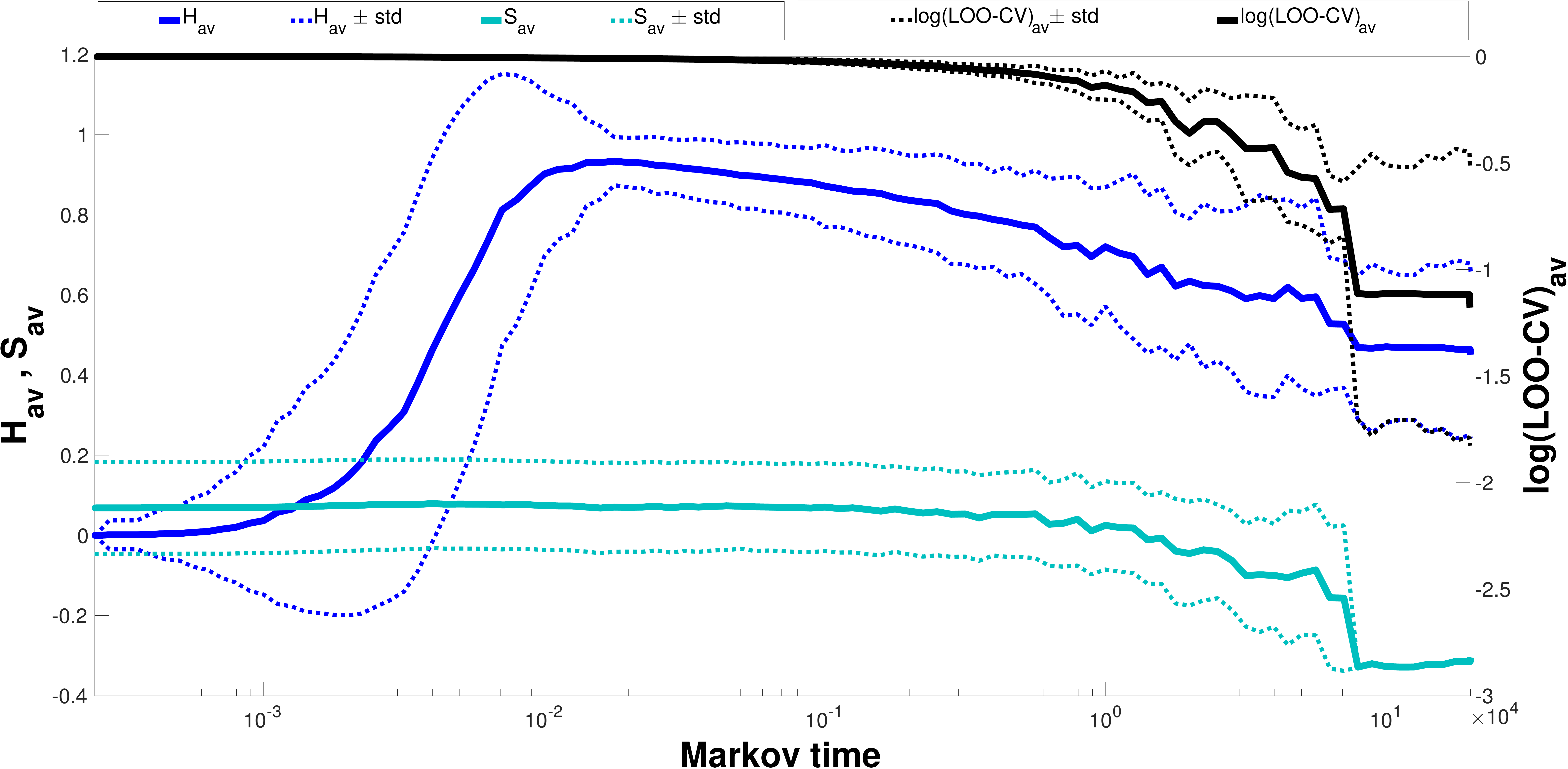}\\
\caption{Clustering performance under heat-shock from a temperature of 25$^{\circ}$C to 37$^{\circ}C$ (HS25-37). Performance under linear functions, average homogeneity ($H_{av}$) and separation ($S_{av}$) between clusters, and non-linear functions based on Gaussian process regression, leave-one-out cross validation ($log(LOO-CV)_{av}$). std: standard deviation.}
\label{fig:Fig_StabilitySolutionHS25-37_Perform} 
\end{center}
\end{figure*}
% ----------------------
% ----------------------

The details of the clustering solutions found for each stress will not be evaluated here. This will be performed elsewhere, by comparing the solutions obtained against other species'. Instead, we focus on the performance of the protocol with standard undirected and directed co-expression graphs that improve the results.

% ------------------------------------------------------------------------------------------------------------------------------------------
% ------------------------------------------------------------------------------------------------------------------------------------------

\subsubsection{Clustering performance} \label{sec:ClusteringPerformanceSGCA}

The performance of the clustering method evaluated here with information on expression dynamics' local geometry was evaluated by calculating the value, for partitions across all resolutions, of 2 sets of measures focusing on different average aspects (see section~\ref{sec:Clustering performance functions} in Supplementary Information): linear and non-linear consistency between time-dependent expression profiles and gene ontology similarity. The average consistency measures selected for this work were the average intra-cluster homogeneity, $H_{av}$ (see Eq.~\ref{eq:Eq_Hav}), and the average Leave-One-Out Cross Validation, $LOO-CV_{av}$ (see Eq.~\ref{eq:Eq_LOOCV}) based on Gaussian process regression analysis. In addition to these intra-cluster performance scoring functions, we also calculated the divergence between profiles through the linear average inter-cluster separation, $S_{av}$ (see Eq.~\ref{eq:Eq_Sav}).

% ----------------------
% ----------------------
\begin{figure*}[!tb]
\begin{center}
\includegraphics[width=1\textwidth]{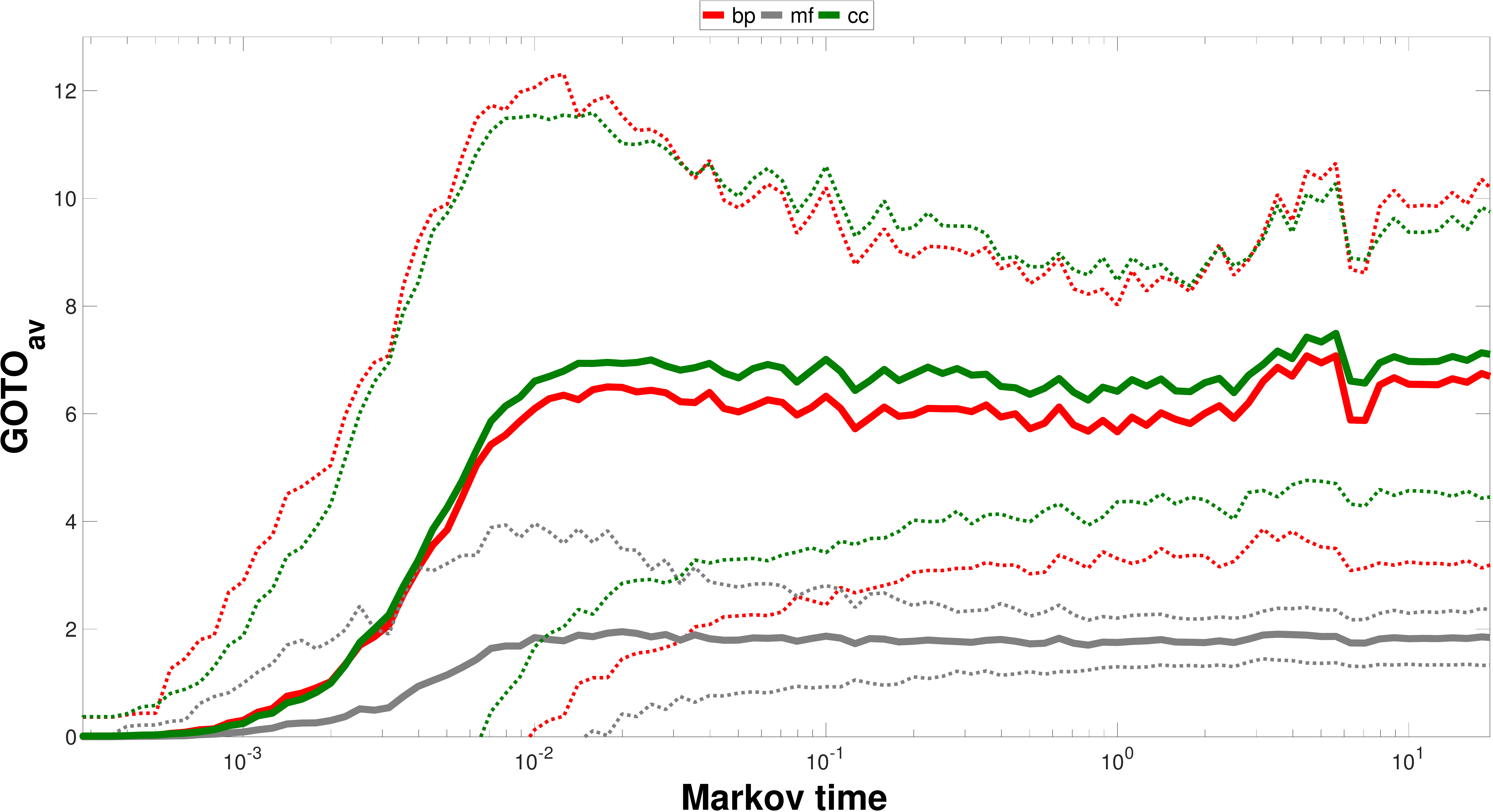}
\caption{Gene ontology clustering performance for heat-shock from a temperature of 25$^{\circ}$C to 37$^{\circ}C$ (HS25-37). Gene ontology term overlap ($GOTO_{av}$) for biological process annotation (bp), molecular function (mf) and cellular component (cc).}
\label{fig:Fig_StabilitySolutionHS25-37_PerformGOTO}
\end{center}
\end{figure*}
% ----------------------
% ----------------------

A good clustering solution should show high homogeneity between elements of the same cluster and, similarly, high predictive probability (intrinsic to $LOO-CV$ \cite{Rasmussen2006}) of any of its elements belonging to the same distribution. On the other hand, a good solution should also reveal low values of $S_{av}$ between elements of different clusters. A semantic similarity function was also used to further evaluate performance. We resorted to  Gene Ontology Term Overlap ($GOTO$, Eq.~\ref{eq:Eq_GOTO})  due to its simplicity and high-fidelity \cite{Datta2006,Mistry2008}. 

The homogeneity and separation profiles (Fig.~\ref{fig:Fig_StabilitySolutionHS25-37_Perform}) indicate that after an initial period where performance is poor (low $H_{av}$ and $S_{av}$ close to zero), the algorithm starts generating partitions that attain a better grouping of expression profiles. The maximum of performance according to the homogeneity evaluation functions is attained after the Markov stability algorithm reaches a point where no singletons exist (see the number of communities or clusters scatter plot in gray in Fig.~\ref{fig:Fig_StabilitySolutionHS25-37}). Before this point, due to the fact that the average homogeneity is a pair-wise performance function, it associates a value of $0$ to each singleton, thus resulting in lower performances. For the $GOTO_{av}$ function similar observations hold (see Fig.~\ref{fig:Fig_StabilitySolutionHS25-37_PerformGOTO} for a typical profile). Concerning the $LOOCV_{av}$ performance function (see Eq.~\ref{eq:Eq_LOOCV} and Fig.~\ref{fig:Fig_StabilitySolutionHS25-37_Perform}), it computes a score for each cluster by eliminating the value of each time-point and computing the marginal predictive probability for the eliminated point. Therefore, even if we have a singleton, a score is calculated, which clarifies the fact that its profile with Markov time differs from that of $H_{av}$ and $GOTO_{av}$ (bp,cc or mf).

% ----------------------
% ----------------------

\begin{table*}[!tb]
\centering
\begin{tabular}{c|c|c|c}
Function        &   Cosine - GP    & BHC-Cosine & BHC-GP\\
\hline
$H_{av}$        &  0.44 - { \bf 0.56} & 0.03 - {\bf 0.97} & 0.18 - {\bf 0.82}\\

$S_{av}$        &  {\bf 0.95} -  0.05  &   {\bf 0.92} - 0.08 & {\bf 0.90} - 0.1\\

$LOOCV_{av}$    &   0.37 - {\bf 0.63}  &   0.41 - {\bf 0.59} & 0.42 - {\bf 0.58}\\

$GOTObp_{av}$   &   0.43 - {\bf 0.57} & 0.33 - {\bf 0.67} & 0.22 - {\bf 0.78}\\

$GOTOcc_{av}$   &   0.46 - {\bf 0.54} & 0.26 - {\bf 0.74} & 0.20 - {\bf 0.80}\\

$GOTOmf_{av}$   &   0.46 - {\bf 0.54} &  0.34 - {\bf 0.66} & 0.22 - {\bf 0.78}

\end{tabular}
\caption{\label{tab:Performance of Markov stability clustering analysis (undirected networks)} Clustering performance (undirected co-expression networks) across distance functions and against Bayesian hierarchical clustering \cite{Cooke2011}. The results presented here are the proportions of clustering solutions where a specific distance function, of each pair, induces better results according to the selected performance measures. Cases that do not add up to 1 exclude instances where performance was equal.}
\end{table*}

% ----------------------
% ----------------------

Regarding the performance of each of the dissimilarity functions, results for both of the original undirected networks as well as an enhancement of edges with causal information are reported below. We also compute the performance against a state of the art algorithm for time-series analysis based on both Gaussian process regression and Dirichlet process mixture (DPM) models, namely the Bayesian Hierarchical Clustering algorithm (BHC) \cite{Cooke2011}. The BHC algorithm was shown to surpass other similar methods in previous studies \cite{Cooke2011} of a number of datasets and, therefore, constitutes the benchmark.

The comparison between dissimilarity function performances was evaluated as follows: for the set of solutions found across all resolutions for each stress, we extracted all that had a partition with a unique number of clusters. For the plateaus, the solution with the lowest partition pair-wise variation of information, found by starting at several initial Louvain conditions, was selected as the most reliable. Subsequently, spline curves were fitted to the value of each of the performance measures as a function of the number of clusters in each partition. The number of times across all resolutions that a particular option performed better, across all stresses, was used as an indication of the performance of one option over the other.

The BHC solutions used for the performance analysis were extracted from the hierarchical dendrogram generated by the BHC package by cutting it at several heights: from those resulting in partitions with 2 up to a maximum value corresponding roughly to a partition with $\frac{N_{genes}}{2}$ clusters. The algorithm used for this task was a purpose written set of routines in R using standard methods readily available from the Bioconductor repository ($https://www.bioconductor.org$). Once again, spline curves were fitted to the value of each of the performance measures as a function of the number of clusters in order to compare performances.

Markov stability analysis always performs better (across all performance measures with exception of $S_{av}$) when the $GP$ distance function was employed (see Table~\ref{tab:Performance of Markov stability clustering analysis (undirected networks)}). In fact, stability analysis either is comparable or surpasses the BHC algorithm with respect to $H_{av}$ and $LOO-CV_{av}$, as well as the gene ontology similarity measures $GOTO_{bp,mf,cc}$, for molecular function, cellular component and biological processes terms. 

The BHC algorithm relies on a function determining at which height the best solution should be found: the posterior probability for a specific branch \cite{Cooke2011} (a weighted version of Eq.~\ref{eq:Eq_GP}). This probability is calculated for each merging branch and the cutting height transformed into a collection of cutting heights. If the value reaches 0.5, than BHC dictates that we are in a scenario where the probability of association of  gene expression profiles starts being lower than that found by mere chance \cite{Cooke2011}. Despite not relying on the same principles as the BHC, the posterior of each partition found according to the Markov stability analysis can also be calculated through the same ratio as BHC does, but in our case with respect to the previous Markov time. This would give us a sense of consistency, under a Bayesian framework, of the progression of the agglomeration process underlying Markov stability analysis. Due to computational restrictions we will not compute it here. Yet, we can evaluate at which resolution stability analysis generates solutions that resemble the best partitions proposed by the BHC and the posterior probability criterion. Overall, we would have to look at very high resolutions in the multi-scale solution determined by Markov stability in order to generate a similar number of clusters, with equivalent allocation of nodes, as the BHC does (see Supplementary Fig.~\ref{fig:DistancetoBHC_allstresses_allstresses}).

% ----------------------
% ----------------------

\begin{table*}[!tb]
\centering
\begin{tabular}{c|c|c}

Function        &   Cosine (undir.) - Cosine (dir.)   & GP (undir.) - GP (dir.)\\
\hline
$H_{av}$        &   0.32 - {\bf 0.68} & {\bf 0.59} - 0.41 \\

$S_{av}$        &   0.04 - {\bf 0.96} & 0.04 - {\bf 0.96} \\

$LOOCV_{av}$    &   {\bf 0.83} - 0.17 & 0.46 - {\bf 0.54} \\

$GOTObp_{av}$   &   0.34 - {\bf 0.66} & 0.49 - {\bf 0.51} \\

$GOTOcc_{av}$   &   0.27 - {\bf 0.73} & 0.49 - {\bf 0.51} \\

$GOTOmf_{av}$   &   0.27 - {\bf 0.73} & 0.49 - {\bf 0.51}

\end{tabular} 
\caption{\label{tab:Performance of Markov stability clustering analysis with undirected and directed networks} Performance of clustering analysis, undirected vs directed, for both cosine and Gaussian process regression dependent distance functions. The results presented here are the proportions of clustering solutions where a specific distance function, of each pair, induces better results according to the selected performance measures. Cases that do not add up to 1 exclude instances where performance was equal.}
\end{table*}

% ----------------------
% ----------------------

%--------------------------------------------------------------------------------------------------------------------------------------------
% ------------------------------------------------------------------------------------------------------------------------------------------

\subsubsection{Enhancing co-expression networks with causality improves performance}\label{sec:RMSTplusCSI}

Causal network identification has been extensively explored in the literature of gene network construction from time-course data \cite{Penfold2011,Bar-Joseph2012}. The approaches range from Bayesian and Dynamic Bayesian networks, to non-parametric methods tailored to non-linear dynamics and Granger causality based methodologies \cite{Penfold2011}. Other areas are also ripe with methods stemming from the Granger causality paradigm \cite{Barnett2014}, which augmented by the Bayesian formalism stands as a competitive alternative. 

In this section, we aim at using aspects related to signal shape and timing captured by this type of approach, in order to enrich the underlying co-expression network of each dataset with causality. Therefore,  we transform the original structures into directed weighted graphs. The use of directed networks under a flow-based community detection paradigm has been reported to increase specificity in the clusters identified, which adds a more refined interpretation of the underlying data \cite{Beguerisse-Diaz2013a,billeh2014}. We will evaluate the performance of this approach by comparing it with the respective undirected cases. We must emphasize that the costs of creating, for example, an original distance matrix based on $GP$ dissimilarity, plus the computation of causality on each remaining edge, are very high. Nevertheless, the intent is not providing an alternative to other algorithms dealing with causality, but enhancing our analysis. On the other hand, if for example we identify a local RMST structure with a linear distance function, e.g. the cosine dissimilarity, which is very fast to compute, the advantages of combining this initial step with the strength of causal structure identification algorithms such as those presented below are promising.

In order to determine the causal structure of each of the co-expression graphs, we relied on the calculation of the posterior over all possible causal structures allowed, $P(M)$ \cite{Penfold2012} (see Methods for details).

% ----------------------
% ----------------------
\begin{eqnarray}
 P(M)=\prod^{|E^{RMST}|}_{i=1}P({\it Pa(i)}|\mathbf{g_{i}},\mathbf{[g_{-i}]},\theta_{i})\label{eq:PosteriorCSI}
\end{eqnarray}
% ----------------------
% ----------------------

Adding flow directionality improves the performance of stability analysis, across most performance measures, for each distance function with respect to the undirected case (see Table ~\ref{tab:Performance of Markov stability clustering analysis with undirected and directed networks}). Regarding the linear cosine dissimilarity only the $LOOCV_{av}$ is not improved. The non-linear $GP$ distance function, on the other hand, succeeds in improving performance across all measures with exception of average homogeneity. 

If, on the other hand, the weights matrix is determined by the algorithm relying on Eq.~\ref{eq:Eq_CSIParentPosterior}, the performance is not improved as much for cosine dissimilarity. Yet, for the GP dissimilarity derived graphs, Markov stability outperforms in 8 out of 9 measures (see also Supplementary Tab.~\ref{tab:PerformanceRMSTplusCSI} for results pertaining to additional weighting alternatives). 

Once again, we must add that the computational costs of creating directed networks, via methods such as those presented in \cite{Penfold2012}, for the 2000 gene set selected for analysis in our work, especially if we compute them for all of the 11 stresses, is very high. Therefore, turning to the RMST algorithm as a way of accelerating the calculations is, in fact, a viable option. 

% ------------------------------------------------------------------------------------------------------------------------------------------
% ------------------------------------------------------------------------------------------------------------------------------------------

\subsection{Network comparison from community structure at multiple resolutions} \label{sec:Network comparison}

Comparison of network structures at multiple resolutions has attracted considerable interest recently \cite{Onnela2012,Lee2014}. The mesoscale can highlight aspects of common organization between structures stemming from completely different sources, effectively allowing for a taxonomy of networks to be determined \cite{Onnela2012}, even if the source of each network is completely different. The flow-based clustering analysis reported above provided an understanding of network structure meso-scale. We can further the study of stress-induced co-expression networks by relating the sequence of sub-structures to other stresses, thus identifying cross-stress features in the process.

Each of the clustering solutions obtained via Markov stability analysis of each of the stresses, in expression space, was further compared to others by resorting to information based similarity functions \cite{Meila2007,Vinh2010}. These have been proven to be very efficient in evaluating differences between partitions and have been extensively explored in the literature as a strong alternative to the adjusted Rand index \cite{Rand1971,Fritsch2009}.

We applied once more the RMST algorithm. Yet, this time it was done in partition solution space with a Mutual Information function adjusted for chance ($AMI$), under an hyper-geometric model for randomness \cite{Vinh2010}. Only cross-stress distances were used. Had we applied the algorithm to the full matrix, i.e. also including intra-stress entries, the resulting partition space network would not have retained a significant amount of cross-stress edges, thus resulting in a star like structure with almost fully connected nodes corresponding to clustering solutions in expression space with less than 6 sub-groups. Additional tests were also done with a normalized version of a variation of information function \cite{Meila2007}, but these showed that $AMI$ is better suited for our case.

As the $GP$ dissimilarity function outperformed the cosine dissimilarity, we chose to focus on the solutions provided by the former. The resulting partition space RMST mostly connects close partitions with similar numbers of communities (Supplementary Fig.~\ref{fig:Fig_Part_RMST_N}). Although the analysis can be performed for all solutions, let us focus on nodes representing no more than $N=50$ clusters. This constitutes a manageable size for detailed analysis and visualization. The sub-graph connecting this subset of solutions is represented in Fig.~\ref{fig:Fig_Part_RMST_Nupto50}. This sub-structure highlights continuity in cross-stress partition space and can be used for extracting a joint cross-stress partition (see section~\ref{sec:Joint_Partition_Selection}), as well as clustering stress-induced co-expression networks, and hence stresses, based on their multi-scale community organization. 

% ------------------------------------------------------------------------------------------------------------------------------------------
% ------------------------------------------------------------------------------------------------------------------------------------------

\subsubsection{Clustering of co-expression networks}\label{sec:Clust of co-expr nets}

\begin{figure*}[!tb]
\includegraphics[width=1\textwidth]{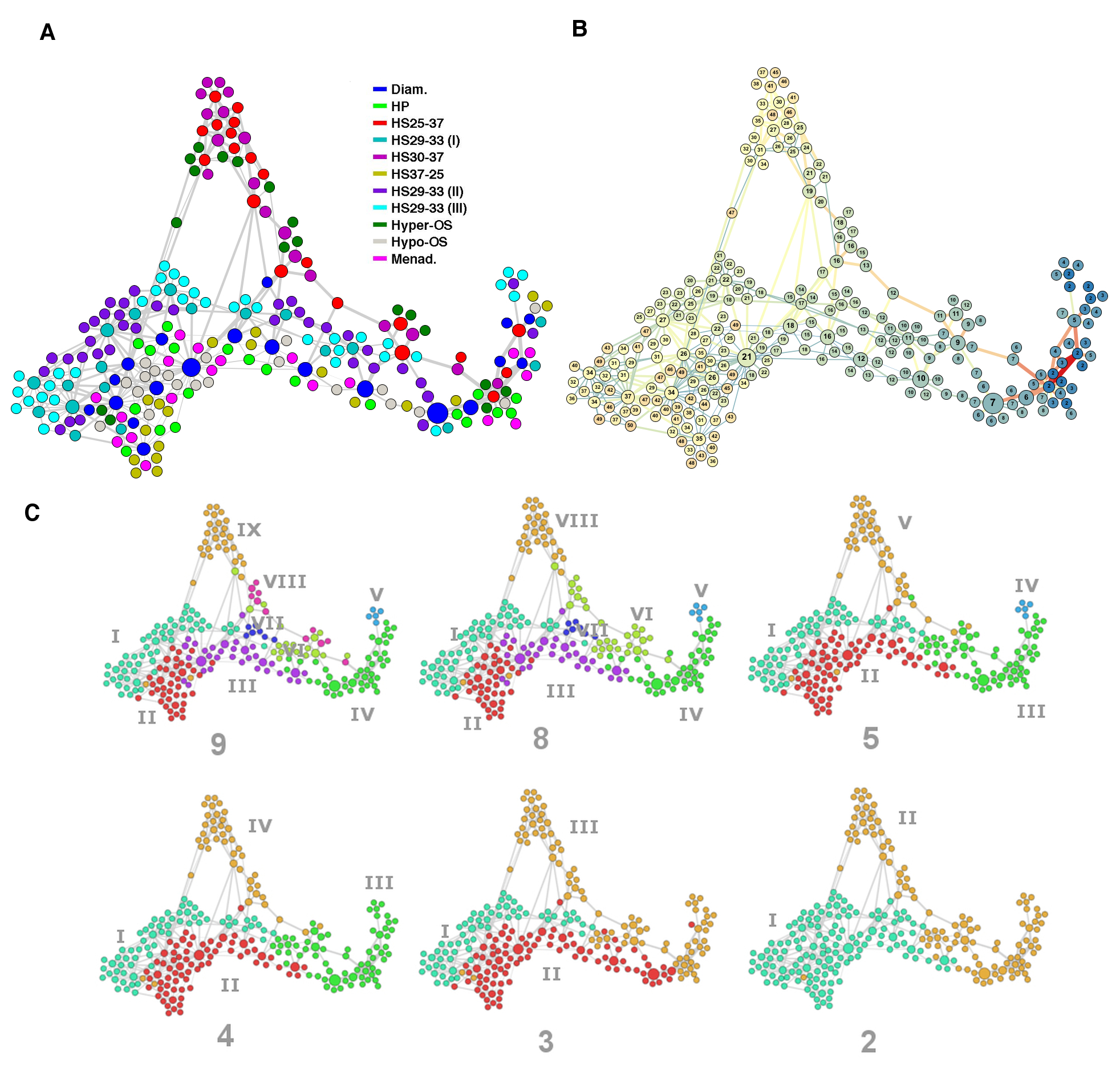}
\caption{\label{fig:Fig_Part_RMST_Nupto50} Detail of network of partitions with up to 50 clusters. (A) Network with nodes coloured by stress. (B) Network with nodes coloured by the number of clusters in the corresponding expression space partition. Detail in Supplementary Fig.~\ref{fig:Fig_Part_RMST_Nupto50_ZOOM}. (C) Community structure of network of partitions with up to 50 clusters. Spatial embedding with force based algorithm in Gephi. Edge thickness is proportional to weight, in this case the $AMI$.}
\end{figure*}

We performed this study on the partition space co-partition network shown in Supplementary Fig.~\ref{fig:Fig_Part_RMST_N} (B), representing connections between nodes corresponding to no more than 50 clusters. A different spatial embedding with nodes identified by stress can also be seen in Fig.~\ref{fig:Fig_Part_RMST_Nupto50} (A).

We chose to resort to Markov stability once more due to its flexibility and efficiency. Other approaches could have been employed that would have imposed a hierarchical sequence of solutions, e.g. the diffusive Shi-Malik algorithm \cite{Shi2000,Delvenne2010}. 

A selected set of solutions, following similar criteria as those used above during the analysis of expression space, can be visualized in Fig.~\ref{fig:Fig_Part_RMST_Nupto50} (C). The solution, for example when 4 clusters arise, seems to group nodes with $N$ of the same magnitude but distinct category of stresses. Effectively, we see that cluster $I$ is associated with nodes belonging to stresses $HS29-33 (I)$, $HS29-33 (II)$ and $HS29-33 (III)$ with $N$ between 14 and 47, thus highlighting the similarity between the multi-scale organization of any mild temperature stress, at higher levels of resolution, applied individually or in conjunction with osmotic stress. Cluster $II$ groups stresses $Diam.$, $HP.$, $Menad.$, $Hypo-OS$, as well as $HS37-25$ with $N$ between 10 and 45, and thus mostly brings together oxidative stresses and stresses with a pattern that is associated with return to resting state, i.e. absence of stress. Cluster $IV$ focuses on $HS30-37$, $Hyper-OS$ and $HS25-37$ with cluster sizes from 11 to 45, which represent stresses with the highest amplitude. Lastly, cluster $II$ encompasses clustering solutions across all stresses with up to 12 clusters, thus representing in most part features stemming from grouping, in the original expression state, a large number of genes, loosing in the process specificity associated with each response.  Interestingly, the solution where nodes are grouped into two communities, separates roughly, between the group of mild heat shocks, oxidative stress and removal of stress agent ($I$), and the group of osmotic and large temperature shock. This had not been seen in the original dendrogram based solely on Jaccard distances between stress-induced networks under a $GP$ dissimilarity (Supplementary Fig.~\ref{fig:Fig_Hclust_Jaccard} (B)).

% ----------------------
% ----------------------

% ----------------------
% ----------------------

% ------------------------------------------------------------------------------------------------------------------------------------------
% ------------------------------------------------------------------------------------------------------------------------------------------

\subsubsection{Multi-resolution joint partition solution across co-expression networks} \label{sec:Joint_Partition_Selection}

Identification of a possible joint partition stemming from the graph in partition space is closely related to a recently proposed algorithm \cite{Kirk2012}; it allows for identification of genes that are clustered together in different types of datasets. Other approaches also relying on cross-dataset features have also proven to be efficient candidates \cite{Heard2005}. As in \cite{Kirk2012}, the different datasets may be compared due to the fact that our algorithm selects the final solution by comparing partitions. Therefore, as was stated before, even if the data is of different sources or it is not immediately comparable due to experimental noise, we are capable of generating a possible joint clustering solution across datasets. Unlike the work presented in \cite{Kirk2012}, our method does not induce a partition in each stress-induced co-expression network, by using a cross-stress score for genes clustering together, while selecting a joint-partition across stresses. It selects the partitions at each resolution and for each stress, found independently, that allow for continuity in the space of partitions, thus establishing a link between stresses. This approach can also be used for comparing expression dynamics under stress across species as long as we focus on the orthologue genes/nodes (explored elsewhere). As in \cite{Kirk2012}, other types of information summarized as a graph, may it be protein-protein interaction, transcription factor-target gene regulatory networks, or gene ontology similarity networks between genes \cite{Yu2010}, can also be incorporated into the analysis. Another approach dealing with cross-network community properties closely related to Markov stability analysis was presented in \cite{Mucha2010}, where a cross-dataset relationship between partitions was imposed by multiplex networks. This is akin to allowing, in our case, for cross-stress diffusion by defining a distance measure between the expression patterns of genes in two different datasets. This would be possible if, in fact, the datasets were comparable with respect to absolute values of expression, which we do not assume.

In order to select a joint partition across networks and resolutions, we need to focus on the most robust solutions. Therefore, for each stress, from the solutions or nodes with the same number of clusters we selected only one showing the lowest average variation of information (see Supplementary Figs.~\ref{fig:Fig_Part_RMST_Nupto50_VI_S} (A) and \ref{fig:Fig_Part_proj_Nupto50_VIbest}). From the remaining connected nodes we have a joint partition that represents continuity in the organization of expression patterns across multiple stresses. We can also further narrow our view by selecting the sub-graph that connects all nodes with degree bigger than 2, and in the process removing tree-like properties of the remaining structure (Supplementary Fig.~\ref{fig:Fig_Part_proj_Nupto50_VIbest}).

The stress that has partitions with the largest degree in the multi-scale joint partition solution is diamide. As was observed in the original study \cite{Gasch2000}, diamide has a response that resembles a combination of heat-shock, hydrogen peroxide and menadione. The proximity of diamide to other stresses is also obvious in Supplementary Fig.~\ref{fig:Correlation_WeiMat_allstresses}.  In fact, the organization of the expression pattern characterizing diamide also has characteristics of removal of a stress, e.g. $HS37-25$. Also noteworthy is the fact that the nodes in Supplementary Fig.~\ref{fig:Fig_Part_proj_Nupto50_VIbest} associated with the $HS37-25$ stress connect only to nodes representing diamide. 

As was mentioned before, other sources of information can be incorporated in the multi-resolution analysis of co-expression networks, for example that were collected in physical interaction and genetic interaction databases such as $BioGRID$. This extra information can also be useful in extracting from each stress only partitions or nodes in Supplementary Fig.~\ref{fig:Fig_Part_proj_Nupto50_VIbest} that effectively represent close experimental interactions (see further details in Supplementary Information, Figs.~\ref{fig:Fig_Links_BIOGRIDNet} and \ref{fig:Fig_Part_InterStressBIOGRID_edges}).

% ******************************************************************************************************************************************
% ******************************************************************************************************************************************
% DISCUSSION
% ******************************************************************************************************************************************
% ******************************************************************************************************************************************

\section{Discussion}

Both gene regulatory and signalling networks have evolved to deal with the diversity of environmental stimuli regularly encountered \cite{Lopez-Maury2008}. Effectively, cells reaction to their environment is done through a multitude of strategies \cite{Mitchell2009,Dhar2013}. This is translated into an overall time-dependent expression response covering thousands of genes. In order to get meaningful insights into how biological systems dynamically respond to their environment, aggregation of similar responses characterizing each gene is of fundamental importance.  Clustering analysis has been an excellent approach to reducing this complexity through identification of overall similarity features. Yet, most techniques do not take into account the multi-scale nature of these profiles and over partition the overall response. In this work, we presented a protocol based on graph-theoretical principles, specifically relaxed minimum spanning trees, as well as those stemming from the diffusion properties on graphs, referred to as Markov stability analysis, that allows for fast and efficient time-series analysis of expression dynamics. This combined approach addresses the importance of both local and global pair-wise similarity between genes. Overall, the performance of the method with respect to a state-of-the-art time-series clustering algorithm was shown to be enhanced under two dissimilarity functions, one linear and another based on Gaussian process regression. We also verified that refining the structure of each co-expression network with causality improves the cohesiveness of the community organization solutions of each network at multiple resolutions.  The combination of the partition solutions at different resolutions with methods identifying causality between clusters \cite{Lu2011,Windram2012,Wu2014} should also improve the construction of reduced models, akin to a renormalization of the underlying networks \cite{Radicchi2008}, thus allowing for a reliable transition from micro to mesoscale descriptions of gene expression patterns. 

The use of graph-theoretical techniques has been a fundamental contribution for understanding expression patterns as well as the role of specific genes with respect to specific clusters identified from the data \cite{Horvath2008}. The method explored here also allows for a comparison of information obtained from different experimental sources, possibly not directly comparable, in our case mRNA expression levels, genetic interaction data and protein-protein interaction data. By matching the community organization of each specific network via information based distance functions in partition space, a joint partition could be extracted that highlights common aspects, at several resolutions, shared by all datasets. Combining the efficiency and speed of the graph based techniques applied here with the powerful Bayesian framework \cite{Heard2005,Kirk2012,Windram2012} should reveal further avenues for co-clustering of datasets in systems biology. In this area of integrative biology, multiple sources of information from the various individual components, making up intra and inter-cellular networks, are crucial in the identification of function \cite{Windram2012}. The methodology presented here can also be applied when comparing expression patterns across species and enhance evolutionary analysis of co-expression networks already proven to increase the probability of function identification \cite{Tirosh2011,Kolar2012}. Effectively, a theory of multi-scale network organization evolution across species is lacking, where the emergence of a feature such as the typical environmental stress response could be explained from a myriad of perspectives: orthologue node comparison, graph alignment \cite{Berg2006,Kolar2012}, evolutionary principles of modular gene regulation \cite{rohtua2013}, dynamical regime dependence on network structure (see for example from a different area \cite{Schaub2015}), as well as evolvability shaped by dynamics \cite{Jimenez2015}. This combined approach could potentially link modularity and other network structures to important features such as the role stochasticity \cite{Li2010} and entropy \cite{Sun2010} in biology.

% ******************************************************************************************************************************************
% ******************************************************************************************************************************************
% METHODS
% ******************************************************************************************************************************************
% ******************************************************************************************************************************************

\section{Methods}

\subsection{Data pre-processing} \label{sec:DataPreProc}

The dataset analysed here was retrieved from the {\it Saccharomyces} Genome Database ($http://downloads.yeastgenome.org/expression/\\microarray/Gasch_2000_PMID_11102521/$). The analysis focuses on stresses that had samples collected at multiple time-points and that showed similar time-scales in their responses. The selected sub-set of stresses corresponds to the gene expression response to:

\begin{itemize}
\item Diamide treatment ($Diam.$), collected 5, 10, 20, 30, 40, 50, 60 and 90 mins.
\item Hydrogen-Peroxide ($HP$) at 10, 20, 30, 40, 60, 80, 100, 120 and 160 mins;
\item Heat-shock from a temperature of 25$^{\circ}$C to 37$^{\circ}C$ ($HS25-37$) at 5, 10, 15, 20, 40, 60 and 80 mins;
\item Temperature shift from 37$^{\circ}C$ to 25$^{\circ}$C ($HS37-25$) at 15, 30, 45, 60 and 90 mins;
\item Hyper-osmotic stress ($Hyper-OS$), with sorbitol, sampled at 5, 15, 30, 45, 60 and 90 mins;
\item Hypo-osmotic stress ($Hypo-OS$), or removal of sorbitol, with samples recovered at 5, 15, 30, 45 and 60 mins;
\item Mild heat-shock from 29$^{\circ}$C to 33$^{\circ}C$ ($HS29-33 (I)$), samples collected at 5, 15, 30 and 90 mins;
\item Mild heat shock in the presence of osmostic stress ($HS29-33 (II)$), observed at 5, 15, 30 mins;
\item Reversal of mild heat shock in the presence of osmostic stress ($HS29-33 (III)$) at 5, 15, 30 mins;
\item Medium heat shock ($HS30-37$) observed at at 5, 15, 30 and 60 mins;
\item Menadione exposure ($Menad.$) at 10, 20, 30, 40, 50, 80, 105, 120 and 160 mins. 
\end{itemize}

For further details on concentrations and specific experimental details see the original paper \cite{Gasch2000}.
The data available from the $SGD$ database had been already pre-processed. According to the information available on the respective website, missing values were imputed using the KNN impute algorithm with nearest neighbours number equal to 10, under the euclidean distance function \cite{Hibbs2007}; technical replicates were averaged. The data for each stress was $log2$ transformed with respect to the unstressed initial instant. 

The expression tables corresponding to the stresses highlighted above and available from $SGD$ database did not include the same genes. In order to compare the responses across stresses, the intersection of the gene lists per stress was used in the subsequent analysis protocol (see Fig.~\ref{fig:Flow_diagram}).

\subsection{Workflow for analysis and clustering of co-expression networks} \label{sec:Flowdiagram}

\begin{figure*}[!tb]
\centering
\includegraphics[width=1.2\textwidth,trim={{0.2\textwidth} {0.1\textwidth} {0.1\textwidth} {0.1\textwidth}},clip]{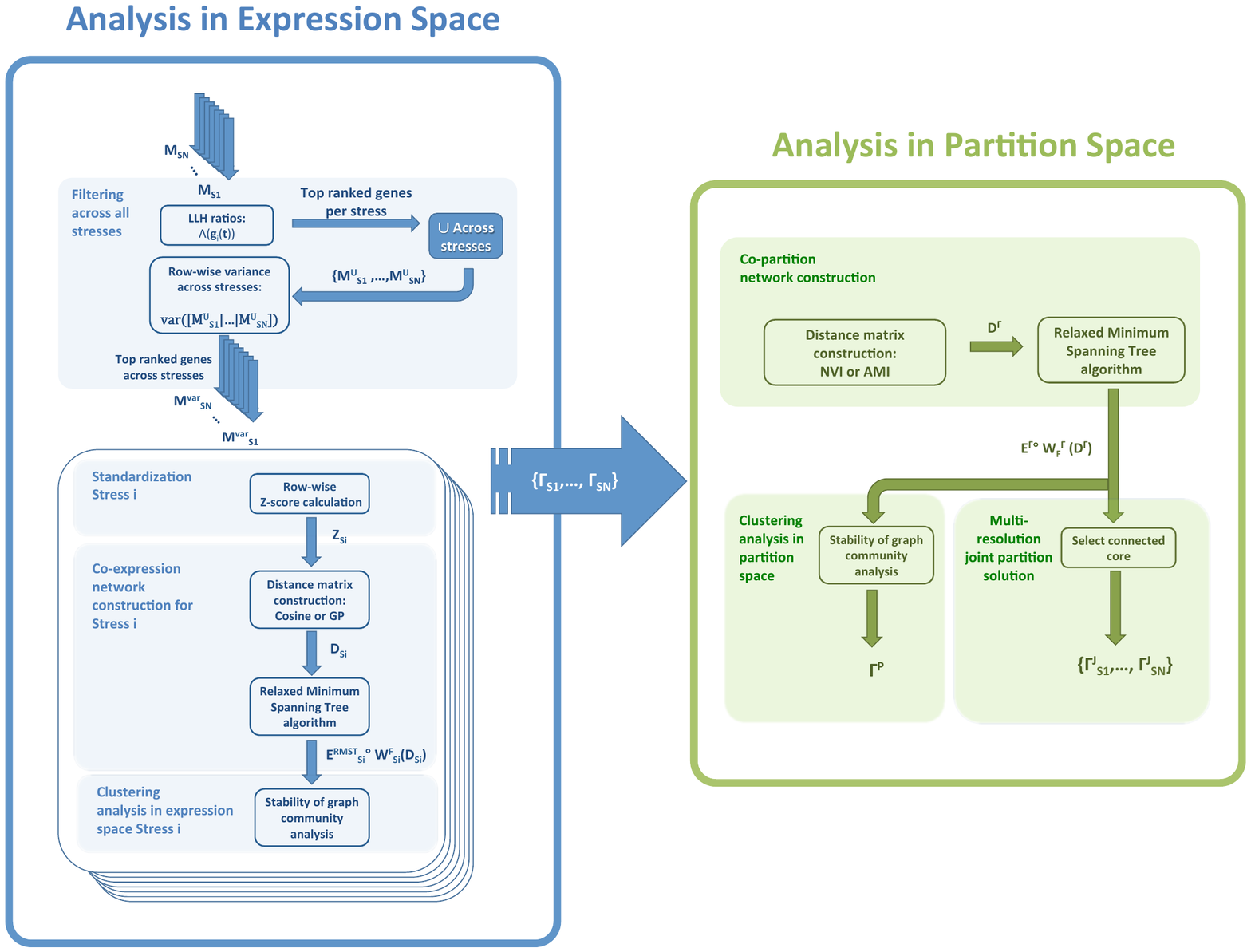}
\caption[Flow diagram for co-expression network analysis]{\label{fig:Flow_diagram} Flow diagram for co-expression network analysis. Blue modules represent analysis in expression space, with an underlying $Cosine$ or $GP$ distance function. Green module represents analysis in partition space with an underlying adjusted mutual information function ($AMI$) \cite{Vinh2010} or a normalized variation of information function ($NVI$) \cite{Meila2007}.}
\end{figure*}

A workflow-diagram representing each stage of the protocol underlying the work reported here can be seen in Fig.~\ref{fig:Flow_diagram}. The blue and green modules involve analysis in expression and partition space, respectively.

%---------------
%---------------
%---------------
%---------------

\subsubsection{Analysis in expression space}

The module that concerns to gene-filtering across all stresses involves a one-sample log-likelihood test for differential expression analysis and selection of the union of the top 1000 ranked genes per stress $Si$, according to Eq.\ref{eq:LLRmain}. The set of stress-specific expression matrices with genes belonging to that union is represented in Fig.~\ref{fig:Flow_diagram} by $\left\{ M^{U}_{S1},...,M^{U}_{SN}\right\}$. The large set of genes produced in the previous step is further reduced by analysing the row-wise variance of the concatenated stress matrices, represented by $var([M^{U}_{S1}|...|M^{U}_{SN}]$; $[..|..|..]$ represents concatenation. After extracting the top 2000 expression profiles that rank the highest according to this new criterion, we further simply the task of expression dynamics analysis under each stress, by mostly focusing on distinguishing features across stresses. While this simplification is a fundamental step, the selected 2000 genes secures that we still maintain a level of common response across stresses that has been proven to be crucial in understanding stress response in yeast \cite{Gasch2000,Gasch2002}. In section~\ref{sec:Overlap_diffexpr_genes}, the overlap between sets of genes in each stress and the final set of selected genes for analysis is reported in detail.

The subsequent module, according to the flow represented by arrows in Fig.~\ref{fig:Flow_diagram}, pertains to graph community analysis of the co-expression networks generated for the set of genes selected in the previous protocol stage. After row-wise normalization, which has been proven to facilitate robust clustering \cite{Golub1999,Cheadle2003}, each stress expression matrix ($Z_{Si}$) is transformed into a distance matrix ($D_{Si}$) by applying the distance functions selected for this work (see section~\ref{sec:Dist_fun}). Construction of the co-expression networks per stress ($E^{RMST}_{Si}$) is performed via the Relaxed Minimum Spanning Tree  (section~\ref{sec:RMST}) and clustering analysis of each of the graphs is conducted by applying Markov stability analysis to $E^{RMST}_{Si} \circ W^{F}_{Si}(D_{Si})$ (see section~\ref{sec:SGCA}). $W^{F}_{Si}(D_{Si})$ represents the full weights matrices for each stress and $E^{RMST}_{Si}$ the undirected and unweighted co-expression graph. From this analysis in expression space we generate a set of stress-induced co-expression networks and respective partitions across multiple resolutions:$\left\{\Gamma^{S1},...,\Gamma^{SN}\right\}$, where $\Gamma^{S1}=\left\{\Gamma_{t_{i}}^{S1},...,\Gamma_{t_{f}}^{S1}\right\}$, with $t_{i,f}$ standing for initial and final Markov times, respectively (see section~\ref{sec:SGCA}).

\subsubsection{Analysis in partition space}

The set of clustering solutions $\left\{\Gamma^{S1},...,\Gamma^{SN}\right\}$ is then evaluated once more by generating a distance matrix ($D^{\Gamma}$) based on the Adjusted Mutual Information ($AMI$, see Eq.~\ref{eq:Eq_AMI}) \cite{Vinh2010}, in this instance in partition space. We also tried the same analysis with the normalized variation of information function ($NVI$) \cite{Meila2007}. The results reported for partition space analysis are, nevertheless, mostly relying on $AMI$. The next step relies once again on the construction of a relaxed minimum spanning tree ($E^{\Gamma}$), this time in partition space (green module in Fig.~\ref{fig:Flow_diagram}). The clustering analysis of $E^{\Gamma}$ allows us to understand how the partitions at each resolution, and in each stress, relate to other stresses'. Therefore,  this analysis gives us a continuous transition between partitions across stresses along a manifold which may inform on the best joint stress clustering solution. In order to generate a clustering solution of stress partitions ($\Gamma^{P}$), we applied once more Markov stability to the matrix $E^{\Gamma} \circ W_{F}^{\Gamma}(D^{\Gamma})$, where $W_{F}^{\Gamma}$ is the full weights matrix derived from the distance matrix $D^{\Gamma}=1-AMI^{\Gamma}$, where $AMI^{\Gamma}=[AMI^{\Gamma}(\Gamma_{t}^{Si},\Gamma_{t'}^{Sj})]$.

A series of other algorithm are available for this type of analysis which are tailored to community detection in graphs, an example being the Shi-Malik algorithm \cite{Shi2000} which can, in fact, be combined with Markov stability analysis (see \cite{Delvenne2010} for details). Here, we decided to resort once more to stability analysis simply due to its versatility and efficiency in different circumstances.

The cross-stress joint partition selection is performed by focusing on a narrow set of solutions connected by the partition space minimum spanning tree, with the highest robustness measured by the average variation of information (see section~\ref{sec:SGCA}). By identifying the connected core in the partition space graph, we arrived at the joint partition solution $\left\{\Gamma^{J}_{S1},...,\Gamma^{J}_{SN}\right\}$. Further integration with datasets from other sources, specifically physical and genetic interaction, is also explored in section~\ref{sec:BIOGRID}.

%--------------------------------------------------------------------------------------------------------------------------------------------
%--------------------------------------------------------------------------------------------------------------------------------------------

\subsection{Filtering of genes for multi-scale network analysis}\label{sec:Filtering}

We focused the analysis of each stress on the same subset of 2000 genes selected according to a one-sample likelihood ratio test \cite{Kalaitzis2011} based on Gaussian process regression (Eq.~\ref{eq:LLRmain}), and the variance of the amplitude of the normalized signals across stresses (see sections~\ref{sec:diff_analysis} and~\ref{sec:Overlap_diffexpr_genes} on differential expression analysis and the thorough analysis of the selected gene sets).

% ----------------------
% ----------------------
\begin{equation}
\Lambda(\mathbf{g}_{i}\mathbf{(t)})=ln(\frac{\argmax_{\theta_{2}} p(\mathbf{g}_{i} \mathbf{(t)} | \mathbf{t},\theta_{2})}{\argmax_{\theta_{1}} p(\mathbf{g}_{i}\mathbf{(t)} | \mathbf{t},\theta_{1})})\label{eq:LLRmain}
\end{equation}

% ----------------------
% ----------------------

In Eq.~\ref{eq:LLRmain} $\mathbf{g}_{i}(\mathbf{t})=(g_{i}(t_{1}),...,g_{i}(t_{N}))$ and  $\theta$ represents the set of hyper-parameters of a specific covariance function (see Eq.~\ref{eq:covariance_kernel} in Supplementary Information). The set of parameters represented by $\theta_{2}$ corresponds to a model where structure in the signal is modelled, whereas the set $\theta_{1}$ assumes that the variation observed in the samples collected across time-points is simply explained by noise and no overall tendency across time that would be associated with differential response to stress.

 The use of Gaussian process regression in differential expression analysis, as is the case of the one-sample likelihood ratio test \cite{Kalaitzis2011}, has been shown to be a more efficient indicator of structure in expression dynamics and, subsequently, in aiding selection of genes that show considerable stress induced activity \cite{Stegle2010,Kalaitzis2011,Windram2012}. The size of the selected set of differentially expressed genes assures that we focus on the most important features of the data for each stress and across stresses. It also secures that we include genes (approximately 1000) that were seen in previous studies to be associated with global responses to stress, namely those identified as belonging to the ESR set of  genes \cite{Gasch2000,Gasch2002,Gasch2007} (see section).

 % ------------------------------------------------------------------------------------------------------------------------------------------
% ------------------------------------------------------------------------------------------------------------------------------------------

\subsection{Dissimilarity functions}\label{sec:Dist_fun}

As was outlined above in section~\ref{sec:Flowdiagram}, the protocol used here involves a two stage step. First, we perform the analysis of each stress independently by resorting to 2 distance function described below. Subsequently, in order to evaluate cross-stress features captured by co-expression graphs, we apply information theoretic functions as a way of capturing distances between organization patterns.

\subsubsection{Expression space}
Any clustering algorithm for time-course data needs to accommodate information about time-dependent profiles that may be absolutely critical as a distinguishing feature \cite{Song2012}. Since linear distance functions such as cosine dissimilarity (Eq.~\ref{eq:Eq_cos}) do not take into account time as a distinguishing factor, i.e. the order of events disappears, it was imperative to find an alternative that would be capable to include nuanced time-dependent details. We opted for a distance function relying on Gaussian process regression (see Eq.~\ref{eq:Eq_GP}) that by default takes time into account through the use of a covariance kernel \cite{Rasmussen2006,Stegle2010,Kalaitzis2011,Cooke2011,Windram2012}. We should underline that functions that take into account time and the underlying sequence of events have been proven to perform better than linear ones \cite{Song2012}.

% ----------------------
% ----------------------

\begin{eqnarray}
d^{cos}_{ij}=1-\frac{ \mathbf{g}_{i}(\mathbf{t})\mathbf{g}^{T}_{j}(\mathbf{t})}{\left\| \mathbf{g}_{i}(\mathbf{t})\right\|_{2}\left\| \mathbf{g}_{j}(\mathbf{t})\right\|_{2}}\label{eq:Eq_cos}
\end{eqnarray}

% ----------------------
% ----------------------

\begin{equation}
\begin{split}
d^{GP}_{ij}&= 1-\\
&\frac{p([\mathbf{g}_{i}(\mathbf{t})|\mathbf{g}_{j}(\mathbf{t})]_{J}|\mathbf{t},{\bf \theta_{J}})}{p([\mathbf{g}_{i}(\mathbf{t})|\mathbf{g}_{j}(\mathbf{t})]_{J}|\mathbf{t},\mathbf{\theta_{J}})+ p(\mathbf{g}_{i}(\mathbf{t})|\mathbf{t},\mathbf{\theta_{i}})p(\mathbf{g}_{j}(\mathbf{t})|\mathbf{t},\mathbf{\theta_{j}})}
\end{split} \label{eq:Eq_GP}
\end{equation}
% ----------------------
% ----------------------

The distance functions captured in Eqs.~\ref{eq:Eq_cos} and ~\ref{eq:Eq_GP} can be converted into a weights function to be used in clustering analysis by simply calculating $1-\frac{d^{cos}_{ij}}{2}$, or $1-d^{GP}_{ij}$, since both are normalized to $1$. The likelihoods in Eq.~\ref{eq:Eq_GP} were calculated according to standard formulas from Gaussian process regression (see section on differential expression analysis in Supplementary Information).

The probability ratio underlying the Gaussian Process regression inspired dissimilarity function (see Eq.~\ref{eq:Eq_GP}) also forms the basis of the Bayesian Hierarchical Clustering (BHC) method \cite{Heller2005,Cooke2011}. Contrary to Eq.\ref{eq:Eq_GP}, in the BHC algorithm the ratio is weighted by the probability of a particular merger in the hierarchical tree, $\pi$, and can be calculated from the conjugate Dirichlet prior characteristic of a Dirichlet process mixture (DPM) model \cite{Heller2005,Cooke2011}; in Eq.~\ref{eq:Eq_GP}, in order to get the numerical recipe underlying the BHC method the numerator is substituted by $\pi p([g_{i}(\mathbf{t})|g_{j}(\mathbf{t})]_{J}|\mathbf{t},{\bf \theta_{J}})$ and the competing model by $(1-\pi) p(g_{i}(\mathbf{t})|\mathbf{t},\mathbf{\theta_{i}})p(g_{j}(\mathbf{t})|\mathbf{t},\mathbf{\theta_{j}})$. Consequently, at each iteration $\pi$ reflects the relative mass of the partition where all points are in one cluster vs all other
partitions consistent with the sub-trees. We must also stress that the indices $i$ and $j$, of the corresponding weighted ratio in the BHC algorithm, may be associated with single genes or entire groups of genes. Then, the profiles $g_{i,j}(\mathbf{t})$ have to be substituted by a matrix corresponding to all the time-profiles of the genes in a set of 2 clusters which are being tested for a merger. When using an underlying Gaussian process inspired distance function we are in fact, as with BHC, sampling from the posterior distribution for the DPM model. Yet, instead of weighting each calculation with $\pi$, we resort, in this work, to the stability function for each resolution. Effectively, in our definition of $GP$ distance $\pi=0.5$. 

One advantage of using the Gaussian process inspired distance function over the alternative linear option is the natural capacity of the this probabilistic framework to deal with replicates. The data set used in our work was collected from $SGD$ database, which had only information on averaged expression values over replicated experiments. Yet, explicit inclusion of the variability captured by replicates has been proven to be fundamental in dealing with experimental noise \cite{Stegle2010,Cooke2011}. Although we do not explore this in the work presented here, the additional degree of complexity brought by including replicated experiments could be dealt with immediately without changing any of the subsequent steps in the protocol followed (see Fig.~\ref{fig:Flow_diagram}).

\subsubsection{Partition space}

Regarding the analysis in partition space, the distance function that provided the best results is based on the mutual information ($MI$) between stress-specific partition solutions at different resolutions. Yet, it is corrected for chance under an hyper-geometric model (Eq.~\ref{eq:Eq_AMI}) by the expected value for $MI$ between two partitions, $\mathbb{E}[MI(\Gamma_{t}^{Si},\Gamma_{t'}^{Sj})]$, and the entropy of each partition, $H(\Gamma_{t}^{Si})$ \cite{Vinh2010}.

\begin{equation}
\begin{split}
&AMI(\Gamma_{t}^{Si},\Gamma_{t'}^{Sj})=\\
&\frac{MI(\Gamma_{t}^{Si},\Gamma_{t'}^{Sj})-\mathbb{E}[MI(\Gamma_{t}^{Si},\Gamma_{t'}^{Sj})]}{max(H(\Gamma_{t}^{Si}),H(\Gamma_{t'}^{Sj}))-\mathbb{E}[MI(\Gamma_{t}^{Si},\Gamma_{t'}^{Sj})]}\label{eq:Eq_AMI}
\end{split}
\end{equation} 

% ------------------------------------------------------------------------------------------------------------------------------------------
% ------------------------------------------------------------------------------------------------------------------------------------------

\subsection{Sparsification of co-expression matrices preserving non-redundant features of time-series data} \label{sec:RMST}

The Relaxed Minimum-Spanning Tree (RMST) algorithm \cite{Beguerisse-Diaz2013,Beguerisse-Diaz2014} is a method that relies both on local and global features of the data to recover a co-expression network from the dissimilarity or distance matrix $D=[d_{ij}]$. After defining $D$, according to specific criteria with more or less granularity, the RMST algorithm constructs a network with adjacency matrix $E$ through a sequence of steps which recover some of the manifold identification principles underlying the application of the ISOMAP \cite{Tenenbaum2000}, and the Diffusion Map \cite{Coifman2006,Lafon2006}, but avoids the pitfalls of under-sampling. Unlike the methodologies for manifold reconstruction just mentioned, the construction of the adjacency matrix relies on obtaining a Minimum Spanning Tree (MST) from matrix $D$. The use of a minimum spanning tree as the starting point forces an initial structure that emphasizes the local geometry of high-dimensional data: two points will only be connected if they are part of a path that is part of a structure that minimizes the distance between all points. Yet, MSTs are very sparse and therefore may not be the most faithful representation of datasets that are not uniformly distributed in high-dimensional space. By comparing distances between any pair of points to those along the geodesic path imposed by the MST, we avoid the problem of under-sampling of state space and avoid having disconnected parts of the network. Effectively, the RMST algorithm compares the longest link ($mlink_{ij}$) between two nodes $i$ and $j$ in any path in the MST with the direct distance between the source $i$ and target $j$ in that path, following the heuristic represented by Eq.~\ref{eq:RMSTalgo}. If $mlink_{ij}$ is not considerably smaller the model is not considered to be sufficient to explain the similarity between the selected nodes and a direct link between $i$ and $j$ is added to the starting MST.

% ----------------------
% ----------------------
\begin{eqnarray}
    E^{RMST}_{ij}= 
\begin{cases}
    1,& \text{if} \ \ mlink_{ij}+\gamma(d_{i}+d_{j})>d_{ij}\\
    0,              & \text{otherwise}
\end{cases}\label{eq:RMSTalgo}
 \end{eqnarray}

% ----------------------
% ----------------------

In Eq.~\ref{eq:RMSTalgo} $d_{i}$ represents the distance to the nearest neighbour and $d_{i,j}$ the distance between genes/nodes $i$ and $j$. $\gamma d_i$ approximates the distribution of data around point $i$ in line with the work explored in \cite{Carreira2005}.

The resulting RMST similarity network is an unweighted, undirected graph $E^{RMST}$. By transforming the distance matrix into a weights matrix we define the weighted graph associated with $E^{RMST}$, $G^{RMST}= \lbrace V,E^{RMST},E^{RMST} \circ W^{F}(D) \rbrace$, where two nodes are connected only if their feature vectors are highly similar, regardless of whether they are neighbours in the original graph or not. The RMST network is sparse if the data in $D$ results from a local geometric structure (which the RMST algorithm tries to recover), and is more amenable to analysis using techniques such as the community detection method we describe below.

The RMST algorithm described above is also fundamental in manifold learning and dimensionality reduction \cite{BorislavVangelov2014}. It provides an excellent framework to jointly apply with traditional techniques such as multi-dimensional scaling. Additionally, these techniques can be used in conjunction for filtering analysis of large datasets  with the objective of selecting genes enhancing continuity in a dataset \cite{BorislavVangelov2014}.

The RMST method was applied to each stress individually in order to generate a sparse dissimilarity matrix and preserve information exclusive to each experiment. The analysis could have been performed on a large matrix resulting from concatenating each of the stress expression matrices, as was the case of the original work \cite{Gasch2000,Gasch2002}. Yet, this would result in time-dependent information characteristic of each stress being averaged out due to the sheer length of each concatenated expression vectors, especially if the cosine dissimilarity is used. Furthermore, if the vectors are concatenated, the $GP$ distance function would not be immediately applicable because crucial to this function is Gaussian process regression, which relies on a kernel taking time into account. Although in this situation each response to stress, for a specific gene, could be considered to be a replicated measurement of the same observable under a different condition, if the response is too different, equivalent to having very noisy measurements, the posterior would be too broad and thus would hinder the amount of non-redundant features the RMST algorithm could extract. In order to improve performance and understand stress related specificity, we analyse stresses independently. Cross-stress features were probed differently, in our case in partition space, with a information theory distance function (Eq.~\ref{eq:Eq_AMI}). Yet, even in partition space the RMST tree can be used to identify close relationships between stress-induced networks, with a distance matrix provided by $1-AMI$.

% ------------------------------------------------------------------------------------------------------------------------------------------
% ------------------------------------------------------------------------------------------------------------------------------------------
\subsection{Combination of causal information and sparse co-expression networks}

As was reported in section~\ref{sec:RMSTplusCSI}, adding directionality to the co-expression networks by the method stemming from Eq.~\ref{eq:Eq_CSIParentPosterior} improves the performance of clustering analysis, across most performance measures. 

\begin{equation}
\begin{split}
&P({\it Pa(i)}|\mathbf{g_{i}},\mathbf{[g_{-i}]},\theta)  =\\
&\frac{P(\mathbf{g_{i}}|{\it Pa(i)},\mathbf{[g_{-i}]},\theta_{i}) P({\it Pa(i)}) P(\theta_{i})}{\sum_{pa(i) \in {\it G(i)}}P(\mathbf{g_{i}}|{\it pa(i)},\mathbf{[g_{-i}]},\theta) P({\it pa(i)}) P(\theta)}
\end{split} \label{eq:Eq_CSIParentPosterior}
\end{equation}

In the previous expression, $\mathbf{g_{i}}=(g_{i}(t_{1}),...,g_{i}(t_{f}))$, a vector of observed expression levels of gene i at the instants $t=1,...f$, and $\mathbf{[g_{-i}]}=[(g_{1}(t_{0}),...,g_{1}(t_{f}-1)^{T},...,(g_{i}(t_{0}),...,g_{i}(t_{t_{f}-1})^{T})$ is the matrix of expression values of all other genes connected to $i$ at previous time-points. 

Due to the fact that the sparse co-expression network stemming from applying the RMST algorithm is the starting point, each set {\it pa(i)} in Eq.~\ref{eq:Eq_CSIParentPosterior} is restricted to the in-degree. Correspondingly, {\it Pa(i)} is simply made of all possible combinations of k elements in groups of 1, 2, up to k elements. In the results presented here, we restricted the analysis to groups of at most 2 elements, which allows for possible non-linear causal dependencies related to cooperation being represented. The calculation of the probability in Eq.~\ref{eq:Eq_CSIParentPosterior} depends on the set of parameters jointly represented by $\theta$, which is characteristic of a squared exponential covariance function (Eq.~\ref{eq:covariance_kernel}) in Gaussian process regression \cite{Rasmussen2006}. The optimal hyper-parameters for each case were found through the Expectation Maximization option of the CSI algorithm \cite{Penfold2012} ($http://www2.warwick.ac.uk/fac/sci/systemsbiology/\\research/software/$).

The end result of combining the RMST algorithm with the calculation of the posterior represented in Eq.~\ref{eq:Eq_CSIParentPosterior} is an asymmetric sparse causal structure matrix, $\mathbf{P}$, with two posterior probabilities for each undirected edge of the original graph. In order to determine the final causal relationship on each co-expression graph, we simply selected only one entry from each pair $(P_{ij},P_{ji})$, the one with the highest value; this results in a matrix $\mathbf{P^{c}}$ where if the edge between node $i$ and $j$ was present in the original graph then either $P^{c}_{ij}=0$ or $P^{c}_{ji}=0$, but not both. Subsequently, two sets of simulations were performed: one resorted to the original weights calculated by each of the dissimilarity functions by imposing the causal structure, i.e. $\mathbf{P^{c}}>0$, on the original weighted RMST (section~\ref{sec:RMSTplusCSI}); the other made use of the actual matrix values $\mathbf{P^{c}}$ (section~\ref{sec:PerformanceRMSTplusCSI}) and weighted each edge of the RMST accordingly, thus discarding the weights calculated through the Cosine and GP dissimilarities. This allowed us to further explore directional probability flow on each graph, with an underlying asymmetric weights matrix, by resorting to stability of graph community analysis for directed networks \cite{Beguerisse-Diaz2014,billeh2014}. In Markov stability when the graph selected for analysis is directed, an extra parameter is added to the dynamics, the `teleportation' component,  which ensures that the random walk on the directed graph has no sinks, is aperiodic and recurrent, i.e. ergodic (see \cite{Lambiotte2008,Beguerisse-Diaz2014,billeh2014,Lambiotte2015} for details).

\subsection{Probability flow information for clustering of expression patterns and co-expression networks} \label{sec:SGCA}

The main method used to perform the clustering analysis in this work is based on the identification of sets of sub-graphs or communities that together maximize a quality function known as stability \cite{Delvenne2010,Lambiotte2008}. This method has been very successful in the analysis of structures that have an intrinsic multi-scale organization and are amenable to being represented has networks \cite{Delmotte2011}. Stability of graph communities analysis relies on the definition of a dynamical Markov process on a graph (see Eq.~\ref{eq:Eq_ContMarkovDyn}) and the quantification of its properties at stationarity. The Markov process can be discrete or continuous. Here, we used the continuous version (Eq.~\ref{eq:Eq_ContMarkovDyn}).

% ----------------------
% ----------------------
\begin{eqnarray}
\dot{\mathbf{p}}(t) = -\mathbf{p}(t)[D_{d}^{-1}L]\label{eq:Eq_ContMarkovDyn}
\end{eqnarray}

% ----------------------
% ----------------------

The dynamical process represented in Eq.~\ref{eq:Eq_ContMarkovDyn} defines a Markov chain on the co-expression network, where the probability density at each instant and over all nodes is encapsulated in the  $N \times 1$ $\mathbf{p}(t)$ probability vector. The diffusion across the co-expression network is dependent on $L=D_{d}-A$, the network or graph Laplacian, with $A$ representing the weighted adjacency matrix, in our case a function of the distance matrix, where the strength of a link is larger if two nodes are closer, and $D_{d}=diag(\mathbf{d})$ is the diagonal matrix stemming from $\mathbf{d}=A \mathbf{1}$ ($\mathbf{1}$ represents $N\times1$ vector of ones), the weighted degree of each node. 

In order to quantify the flow across the network we resort to a quality function known as Stability, $r(\Gamma_{t},t)$ (see Eq.~\ref{eq:Eq_stability}), which is the trace of the clustered auto-covariance of the dynamical process on the graph \cite{Delvenne2010,Lambiotte2008}. A community or cluster, under this framework, is identified as relevant at a particular instant, the Markov time t, if the dynamical process taking place has a higher probability of being trapped in its set of nodes than would be expected at stationarity. This allows for the quantification of the quality of partitions at several time-scales of the diffusion process.

% ----------------------
% ----------------------

\begin{equation}
r(\Gamma_{t},t)=trace(\Gamma_{t}^{T}[\Pi P(t)- \mathbf{\pi}^{T} \mathbf{\pi}]\Gamma_{t}) \label{eq:Eq_stability} 
\end{equation}

% ----------------------
% ----------------------

In Eq.~\ref{eq:Eq_stability}, $\Gamma_{t}$ represents the $N \times c$ indicator matrix, where an entry is equal to 1 if node $i$ on the co-expression network belongs to community $j$, with $j \in {1,...,c}$, at Markov time $t$.  For the particular dynamics represented in Eq.~\ref{eq:Eq_ContMarkovDyn}, $\pi$, the stationary distribution of the underlying Markov chain, is given by $\mathbf{\pi}=\frac{\mathbf{d}^{T}}{2m}$, where $m$ is the total weight in the network. In addition, $\Pi=diag(\mathbf{\pi})$ and $P(t)=e^{(-t D_{d}^{-1} L)}$ represents the probability transition matrix.

By finding the partition $\Gamma_{t}$ that maximizes $r(\Gamma_{t},t)$, at each Markov time $t$, we are able to find structure in the data at several time-scales of the diffusion process. Therefore, $t$  acts as a resolution parameter revealing the communities at different scales, from finer, where each community is made of a smaller number of nodes, to coarser, where the number of communities is much lower but each encompasses larger portions of the set of nodes.

Although finding a partition for each Markov time is a NP-hard problem, there are several optimization heuristics for community identification that efficiently reach a solution for our problem. The method we resort to here is the efficient Louvain algorithm \cite{Blondel2008}. Although the Louvain algorithm is deterministic, the optimised solution depends on the initial or seed condition. Here, we find the best solution by starting at a number of 100 initial conditions.

A good partition in Markov stability can be identified by its persistence across a considerable Markov time instants, revealed by a plateau in the number of communities (see for example Fig.~\ref{fig:Fig_StabilitySolutionHS25-37}). In each of the plateaus several options are available that may help in selecting the best partition in the resolution interval where the plateau occurs: consistency between solutions obtained from all the Louvain runs, measured through the average variation of information, and significance of the partitions with respect to a null model \cite{Delmotte2011}. We can also calculate the significance of each of the performance measures used in our work with respect to the same null model derived, for instance, from the original RMST by rewiring the graph \cite{Maslov2002}. Due to time restrictions and the sheer number of runs necessary to evaluate the distribution of measures for each null model graph, here we rely solely on the variation of information \cite{Meila2007} between all the solutions found by starting at several Louvain initial conditions. 

The methodology described above has been extended to directed networks in recent works where a new transition matrix is devised by including a 'teleportation' parameter \cite{Lambiotte2008,Beguerisse-Diaz2014,Lambiotte2015}. Here, we also use stability analysis in directed networks constructed via causal network identification methods.

One important difference between Markov stability analysis with a $GP$ distance function and the BHC is that stability analysis does not impose hierarchy, it arises from the diffusion properties of the dynamical system defined on a graph, here the co-expression relaxed minimum spanning tree found for each stress. Although both algorithms provide invaluable information from the underlying system, one advantage over the BHC algorithm is the execution speed. As with BHC, we start from all possible merging pairs when initially we calculate the distance matrix. Yet, after this initial step, the construction of the co-expression network and the stability analysis across all Markov times is much faster. Even accelerating the calculation of each of the mergers in the BHC algorithm \cite{Darkins2013}, the combination of the RMST algorithm plus stability analysis stands as a computationally less intensive task. Moreover, the Markov time $t$ used in the stability function (see Methods) is in fact a resolution parameter \cite{Delvenne2013}, and the calculation of a partition solution can be computed by starting at any value of $t$, thus giving us flexibility on what resolution we are looking for. This control parameter is currently not a feature available in the BHC analysis.

One option that could be tested is to combine the speed of stability analysis with the underlying formalism of BHC. If we zoom in on the community structure of a specific network by tuning the resolution parameter characteristic of stability, the resulting partition at that resolution can be fed as the seed to the BHC algorithm, and from that point on a hierarchical clustering solution can be imposed, therefore leading to efficient sampling from the posterior clustering solution without resorting to additional randomization features proposed for algorithm acceleration \cite{Darkins2013}. These computational experiments will be reported elsewhere. 

Due to the formalism of DPM models underlying the BHC algorithm, defining a stochastic process on a RMST that would result in an equivalent formalism to that of mixture models could be an invaluable tool for determining community structures on multiplex networks, either consisting of different types of networks \cite{Kirk2012} or networks at different time slices \cite{Mucha2010}.

Merging the two approaches to clustering analysis tested in this work would allow a well rounded perspective of each dataset by combining the strengths of the Bayesian and the graph theoretical algorithms.

% ******************************************************************************************************************************************
% ******************************************************************************************************************************************
% CODE AVAILABILITY
% ******************************************************************************************************************************************
% ******************************************************************************************************************************************

\section{Code availability}

\paragraph*{Differential expression analysis:} The {\it gprege} R package ($http://www.bioconductor.org/\\packages/release/bioc/html/gprege.html$) was used in order to calculate the log-ratio likelihood scores for each gene (see Eq.~\ref{eq:LLRmain}).

\paragraph*{Dissimilarity matrix and co-expression network construction:} The construction of dissimilarity matrices according to the functions outlined above in section~\ref{sec:Dist_fun} can be found here: $https://github.com/nunonene/Multi-scale-analysis-and-clustering-of-stress-induced-co-expression-networks$. For the code used for RMST construction see \cite{Beguerisse-Diaz2013,Beguerisse-Diaz2014,BorislavVangelov2014}.

An implementation of the information based functions for the construction of networks in partition space ($AMI$) can also be downloaded from the respective project page at $https://github.com/nunonene/Multi-scale-analysis-and-clustering-of-stress-induced-co-expression-networks$. Results where the variation of information function was used were found by the script provided with the stability package (see below).

\paragraph*{Clustering performance evaluation} Routines in C++ and Matlab for evaluation of clustering performance and for comparison with Bayesian clustering analysis are also provided at the same address at $https://github.com/nunonene/Multi-scale-analysis-and-clustering-of-stress-induced-co-expression-networks$.  In order to cut the hierarchical dendrograms generated by the BHC algorithm, widely available routines from bio-conductor were used (not provided).

\paragraph*{Bayesian Hierarchical Clustering:} The code used for Bayesian Hierarchical Clustering is available from bio-conductor: $https://www.bioconductor.org/packages/release/bioc/\\html/BHC.html$.

\paragraph*{Clustering analysis:} An efficient code for probability flow-based analysis, used for clustering of gene expression patterns and co-expression networks, which includes the Louvain optimization algorithm, and can be used for both undirected and directed networks, was downloaded from: $github.com/michaelschaub/PartitionStability.$.

\paragraph*{Causality identification on co-expression networks:} Causality or direction on each co-expression network was found with the Expectation Maximization option of the CSI algorithm \cite{Penfold2012} available from $http://www2.warwick.ac.uk/fac/sci/systemsbiology/\\research/software/$.

% ------------------------------------------------------------------------------------------------------------------------------------------
% ------------------------------------------------------------------------------------------------------------------------------------------

\section{Acknowledgements}

N.R.N. acknowledges support from the Combinatorial Response in Stress Pathways (CRISP) consortium, funded by the BBSRC SABR (Systems Approaches to Biological Research) initiative, during the period this project was developed. N.R.N. would also like to thank Professor Barahona and Dr. Vangelov for kindly providing the original code for relaxed minimum spanning tree construction and stability analysis, and for discussions on the applicability of these tools to the analysis of gene expression patterns.

% ------------------------------------------------------------------------------------------------------------------------------------------
% ------------------------------------------------------------------------------------------------------------------------------------------

\section{Author statement}

N.R.N. is solely responsible for the analysis, interpretation of the results and the final text presented in this paper. 

% ------------------------------------------------------------------------------------------------------------------------------------------
% ------------------------------------------------------------------------------------------------------------------------------------------

\section{Additional information}

\textbf{Supporting Information} Further results supporting those reported in the Main Text. All supplementary figures are available at:$https://github.com/nunonene/Multi-scale-analysis-and-clustering-of-stress-induced-co-expression-networks/PaperWithSupplFigs$.

% ------------------------------------------------------------------------------------------------------------------------------------------
% ------------------------------------------------------------------------------------------------------------------------------------------

\bibliographystyle{unsrt} 

\bibliography{SC_NRN2017.bib}

\newpage

% *********************************************************************************************************************************************************************
% Supporting Information
% *********************************************************************************************************************************************************************

\onecolumn

\renewcommand\thefigure{S\arabic{figure}}  
\setcounter{figure}{0}

\renewcommand\theequation{S\arabic{equation}} 
\setcounter{equation}{0} 

\renewcommand\thetable{S\arabic{table}} 
\setcounter{table}{0} 

\renewcommand\thesection{S\arabic{section}} 
\setcounter{section}{0}

\addtocontents{toc}{\protect\setcounter{tocdepth}{3}}
\addtocontents{lof}{\protect\setcounter{tocdepth}{1}}
\addtocontents{lot}{\protect\setcounter{tocdepth}{1}}

{\flushleft\LARGE
\textbf{Supporting Information:}\\[2ex]
\Large
\textbf{Multi-scale analysis and clustering of co-expression networks}
}

{\flushleft
Nuno R. Nen\'{e}$^{1,*}$} 
{\flushleft
${}^1$Department of Genetics, University of Cambridge, Cambridge, UK\\[1ex]
${}^*$nunonene@gmail.com
}

\normalsize
%\singlespacing

\tableofcontents

\newpage

\listoffigures

\newpage 
 
\listoftables

\newpage

% ------------------------------------------------------------------------------------------------------------------------------------------
% ------------------------------------------------------------------------------------------------------------------------------------------

\section{Differential expression analysis} \label{sec:diff_analysis}

The differential expression analysis step was performed by following the method developed in \cite{Kalaitzis2011A}, with genes being ranked according to the log-ratio of marginal likelihoods $\Lambda(\mathbf{g}_{i}\mathbf{(t)})=ln(\frac{\argmax_{\theta_{2}} p(\mathbf{g}_{i} \mathbf{(t)} | \mathbf{t},\theta_{2})}{\argmax_{\theta_{1}} p(\mathbf{g}_{i}\mathbf{(t)} | \mathbf{t},\theta_{1})})$, where $\mathbf{g}_{i}(\mathbf{t})=(g_{i}(t_{1}),...,g_{i}(t_{N}))$ and  $\theta$ represents the set of hyper-parameters of a specific covariance function (see Eq.~\ref{eq:covariance_kernel}), for competing models: one representing signal and other just sampling noise. The marginal likelihood is given by:

\begin{eqnarray} 
p(\mathbf{g}_{i}\mathbf{(t)}|\mathbf{t})=\int p(\mathbf{g}_{i}\mathbf{(t)}|f(\mathbf{t}),\mathbf{t}) p(f(\mathbf{t})|\mathbf{t}) df \label{eq:marginallik_GPR}
\end{eqnarray}

In Eq.~\ref{eq:marginallik_GPR} $f\mathbf{(t)}$ represents an underlying function estimate assumed to be related to the actual observation points $\mathbf{g}_{i}(\mathbf{t})$ through $\mathbf{g}_{i}(\mathbf{t})=f\mathbf{(t)}+ \epsilon$, with $\epsilon$ standing for Gaussian observational noise or error. The prior distribution $p(f\mathbf{(t)}|\mathbf{t})$ over the latent functions $f(\mathbf{t})$ is assumed to be a Gaussian process, $f\mathbf{(t)}|\mathbf{t} \sim GP(m(\mathbf{t}),K_{f}(t_{i},t_{j}))$, where $m(\mathbf{t})=\left\langle f\mathbf{(t)}\right\rangle$ and $K_{t}(t_{i},t_{j})= \left \langle f(t_{i}-m(t_{i}))(f(t_{j}-m(t_{j}) \right \rangle$. Furthermore, $\mathbf{g}_{i}|f(\mathbf{t}) \sim N(f|\sigma_{n}^{2}I)$. The integral in Eq.~\ref{eq:marginallik_GPR} can be evaluated analytically to give the log-marginal likelihood (see Eq.~\ref{eq:marginallik_GPR_sol})\cite{Rasmussen2006A}.

\begin{eqnarray} 
ln(p(\mathbf{g}_{i}(\mathbf{t})|\mathbf{t}))= -\frac{1}{2}\mathbf{g}_{i}(\mathbf{t})^{T}K_{t}^{-1}\mathbf{g}_{i}(\mathbf{t})-\frac{1}{2}ln\left|K_{t}\right|-\frac{N}{2}ln(2\pi)\label{eq:marginallik_GPR_sol}
\end{eqnarray}

where $K_{t}=K_{f}+\sigma^{2}I$. The covariance function $K_{f}$ is chosen to be the squared-exponential kernel (Eq.~\ref{eq:covariance_kernel}).

\begin{eqnarray}
K_{t}=\sigma_{f}^{2}exp(-\frac{(t_{i}-t_{j})^{2}}{2l^{2}})+\sigma_{n}^{2}\delta{ij}\label{eq:covariance_kernel}
\end{eqnarray}

Therefore, the set of parameters $\theta$ are $\sigma_{f}$, $\sigma_{n}$ and the scale factor $l$. The {\it gprege} R package ($http://www.bioconductor.org/packages/release/bioc/html/gprege.html$) was used in order to calculate the log-ratio likelihood scores for each gene. The selected gene set for extensive analysis presented in this work is analysed below in a separate section evaluating the overlap between stress-specific sets.

% ------------------------------------------------------------------------------------------------------------------------------------------
% ------------------------------------------------------------------------------------------------------------------------------------------

\section{Overlap between co-expression network structures by hierarchical clustering based on a normalized Jaccard dissimilarity function}\label{sec:JaccardDistHClust}

As was observed in the main text, each stress co-expression network has a structure that shares a very small amount of edges with any of the other stresses (see Fig.~\ref{fig:Fig_Hclust_Jaccard}). The dendrogram for each dissimilarity function shows, nevertheless, clear groups joining stresses belonging to the same category, e.g. temperature shock. This encourages a more detailed approach to evaluating if the co-expression network structures share other features, despite the average large distance between clusters. Here, the main objective is to perform this by looking at common features associated with community structure.

\section{Detailed analysis of network properties for each stress}\label{sec:AllNetProp}

\subsection{Connectivity based properties of stress induced co-expression RMSTs}\label{sec:ConnProp}

The RMST algorithm relies on identifying local geometry securing that if two genes are connected there is significant information securing proximity between expression profiles (see Methods in main text). Despite one of the additional features of the RMST algorithm being the retention of edges not identified by the original MST but that against which we do not have enough information, it is expected that differences between stresses will be encoded in network structure.

\subsubsection{Stress-specific network connectivity is very different even when original matrices are strongly correlated}  Let us focus for a moment on the stresses $HP$ and $Menad.$, in order to illustrate the differences between two responses captured by network structure. These were reported in the original study by Gasch and coworkers \cite{Gasch2000A} as being largely identical, despite the effect of each of these agents generating different reactive species. There are several layers to this observation that are captured in the process of sparsification via the RMST algorithm. Starting with the correlation between the full weighted matrices generated for each stress, we can observe that the two stresses mentioned above have among the highest correlations between pairs of stresses of supposedly different nature, within the same distance and across distances (Fig.~\ref{fig:Correlation_WeiMat_allstresses}), although not the highest. This indicates, in the first instance, that the response induced by the two stresses is similar at a coarse-grain level. Moreover, although the weights heterogeneity measured by the coefficient of variation is very different for the full matrices of each stress of this pair (see Fig.~\ref{fig:Heterogeneity_allstresses_CV} (A)), the heterogeneity is very similar when the weighted RMSTs are taken into account (Fig.~\ref{fig:Heterogeneity_allstresses_CV} (B)). Once more, the similarity between the stresses in question is reinforced. Yet, specific aspects of each response are captured in the sparsification process and the structures of the respective graphs resulting from the RMST algorithm employed here are significantly different: these two stresses are clustered together when only connectivity is taken into account but not in the first agglomeration step of the dendrogram, for both distance functions (Fig.~\ref{fig:Fig_Hclust_Jaccard}). Therefore, the finer details of the structure, captured by connecting only genes that are locally and globally alike, show dissimilarities that may be an important feature when looking at community structure (addressed in the Main Text) and thus in identifying function of specific sets of genes.

In the case of the diamide expression response, also showing one of the highest graph densities (Fig.~\ref{fig:Fig_RMSTsDensity}), the weights matrix cross-stress correlation is the highest among pairs of different types of stresses (Fig.~\ref{fig:Correlation_WeiMat_allstresses}). In fact, diamide elicits a response which was observed in the original study to be like a composite of heat-shock, hydrogen peroxide and menadione. Also consistent with this is the fact that diamide has the highest pair-wise intersection of genes ranked according to the one-sample likelihood ratio test used for differential expression analysis, both in the original set of genes (Fig.~\ref{fig:Fig_intersect_top1000}), as well as in the 2000 final set of selected genes (Fig.~\ref{fig:Fig_intersect_top1000_inFS}). Nevertheless, as was observed for all pairs of stresses, the diamide co-expression network is still very different from other structures (see Fig.~ \ref{fig:Fig_Hclust_Jaccard}), also confirming the importance of identifying the geometry of the data in order to capture specific stress responses.

\subsubsection{Co-expression network density highlights heterogeneity in original matrices} Generally, the representation of data features via the co-expression graphs arising from the RMST algorithm should follow from the distributed heterogeneity observed in the starting distance matrices. If the data follows a local geometry the networks are sparse, if not they are expected to be denser. Overall, the distance matrix for cosine dissimilarity converted into a matrix of weights reveals that the highest graph densities are those of the stresses that have the lowest dispersion in full matrices and highest in RMSTs (see Tab.~\ref{tab:Spearman correlation all measures}). For $GP$ dissimilarity the negative correlation observed for full matrices is not as pronounced. Contrary to before, the relation between ranks of weights heterogeneity and ranks of density in final RMSTs is exactly the opposite for GP dissimilarity. Even if we remove the stresses with the most contrasting densities, i.e. $HS29-33(II)$ and $HS29-33(III)$, the negative rank correlation is still maintained, albeit smaller. Therefore, we conclude that the information retained when the RMST algorithm is applied with different distance functions is different. Overall, the weights remaining in the RMSTs stemming from the $GP$ dissimilarity are more homogeneous and the graphs represent the non-redundant features of the data with fewer connections, thus pointing to there being a local geometry at play under this function. As is demonstrated in the main text, the GP dissimilarity function generates better clustering performances. 

One striking aspect about all the stresses is that the smallest densities, obtained through the cosine dissimilarity, are associated with combined mild heat-shock and osmotic stress ($HS29-33(II)$ and $HS29-33(III)$). The reverse scenario is observed for GP dissimilarity. These correspond to the stresses where combinatorial-like and serial stress was applied: in $HS29-33(II)$ the cells were grown in a mild temperature environment $29^{\circ}C$ and re-suspended in $33^{\circ}C$ medium but had an additional $1 M$ sorbitol osmotic stress throughout the experiment; in $HS29-33(III)$ the osmotic stress was removed as the temperature stress is increased to $33^{\circ}C$. Curiously, the remaining mild temperature related stresses, $HS29-33(I)$ and $HS30-37$, reveal sparse structure with both distance functions, with the respective graph density being commensurable with the overall trend. The observable differences in the combined stresses highlighted above may stem from a resolution limit in the $GP$ dissimilarity arising, as we understand it, from the number of time-points in the respective expression time-series. The combinatorial and serial stresses were collected only up to 30 mins at 3 different instances, whereas the rest of the stresses were collected for longer at between 4 and 9 different time-points. The reduced number of 3 time-points may not be sufficient information, although this is dependent on signal shape, for performing Gaussian process regression maximizing the difference with respect to noise. Another reason for the result pertaining to mild heat-shock combined with osmotic stress stems from the strict criteria underlying the RMST algorithm, controlled by $\gamma$ (see Methods in main text), for maintaining a link between two nodes. At the level used for co-expression network construction, $\gamma=0.5$, and particularly for the $GP$ dissimilarity function, the evidence against the dense networks observed is clearly not sufficient.

%-------------
%-------------
\begin{table}[ht]
\begin{center}
\begin{tabular}{c|c|c}

Function        &   Cosine & GP\\
\hline 
$\rho_{S}(CV_{RMST},GD)$   &  0.98 (p $<<$ 0.05)  &  -0.64\\

$\rho_{S}(CV_{FULL},GD)$   & -0.85 (p $<<$ 0.05)  &  -0.55\\

$\rho_{S}(CV_{RMST},CPL)$  &  0.98 (p $<<$ 0.05)  &   0.02\\

$\rho_{S}(GD,CPL)$         &  0.98  &   0.34\\

$\rho_{S}(CPL,ESR)$        &  0.50  &  -0.11\\
 
$\rho_{S}(CV_{RMST},ESR)$  &  0.55  &   0.63 (p $<$ 0.05)\\
 
$\rho_{S}(CV_{FULL},ESR)$  & -0.41  &   0.61 (p $<$ 0.05)\\
 
$\rho_{S}(GD,ESR)$         &  0.54  &  -0.32\\
 
\end{tabular}
\caption[Spearman rank correlation, $\rho_{S}$, between graph properties]{\label{tab:Spearman correlation all measures} Spearman rank correlation, $\rho_{S}$, between graph properties. $CV_{RMST (FULL)}$: coefficient of variation for weighted RMSTs (or FULL weighted matrices). GD: graph density. CPL: characteristic path length. ESR: number of environmental stress response core set of genes in top 1000 gene list ranked according to the differential expression analysis likelihood ratio test. p:p-value. For the entries where a p-value is not reported, the result was not significant at a 0.05 significance level.}
\end{center}
\end{table}
%-------------
%-------------

\subsubsection{RMST density correlation with stress intensity is dependent on distance function} Regarding intensity of the stress applied, demonstrated in the group of temperature shocks, i.e. $HS25-37$, $HS29-33(I)$, $HS30-37$ and $HS37-25$, the amplitude of the stress correlates with heterogeneity in the response and the density of the final graphs derived from the cosine distance function ($\rho_{S}=0.95$, for both). For $GP$ dissimilarity a different outcome is observed, $\rho_{S}=0.95$ and $\rho_{S}=-0.74$, respectively, the latter stemming from the negative value of $\rho_{S}(CV_{RMST},GD)$, which equals to $-0.6$ if we restrict to temperature stresses (compare with Tab.~\ref{tab:Spearman correlation all measures}).

\subsubsection{RMST weights heterogeneity correlates with number of ESR genes selected for analysis} Interestingly, there is a rank correlation of $0.55$ and $0.63$, for cosine and GP dissimilarity, respectively, between heterogeneity in the generated co-expression networks and the number of ESR genes belonging to the selected set of genes for RMST construction that, according to the likelihood ration test (see Methods in main text), were in the top 1000 (see also Fig.~\ref{fig:Fig_intersect_top1000_inFS}). If instead of the RMSTs we perform the same calculation with the full weights matrices, the rank correlation coefficient trend is not maintained for cosine dissimilarity. The algorithm under the linear distance function seems to extract features consistent with the correlation reported above from the full matrix. Contrary to this, the $GP$ function imposes from the outset these features on the interaction weights in the full matrix. We must also report that, with both distance functions there are no significant differences between heterogeneity of weights (from a global point of view) belonging to interactions between ESR genes and ESR genes, ESR genes and others, nor between genes not belonging to the ESR set. Therefore, the rank correlation between number of ESR genes in the top 1000 differentially expressed genes and heterogeneity arises simply from the geometric features of the data identified by each distance function.

\subsubsection{Global aspects of stress-induced networks represented in average path lengths} We can also complement the analysis above by considering global aspects of the structures generated for each stress, in our case by calculating the simple average path length. The data plotted in Fig.~\ref{fig:Fig_RMSTsDensity} (B) for the average path length roughly follows the trend of graph density ( $\rho_{S}= 0.98$ for cosine dissimilarity, and $0.34$, for Gaussian process dissimilarity, Tab.~\ref{tab:Spearman correlation all measures}) and thus also consistently points to the significant differences between stresses observed before. These results are not intuitive. Generally we would expect that denser graphs would have smaller average paths lengths. The fact that positive Spearman correlations are observed between graph density and characteristic path length implies an interesting feature of the RMST algorithm: it will allow for a larger proportion of the full matrix to remain with a larger average shortest path between any points, which is consistent with the locality condition governing the algorithm, the trend being more pronounced in the linear distance function. Irrespective of graph density, the characteristic path length reveals aspects of cohesiveness and overall difference between expressions patterns of all genes under different stresses. For all stresses the cosine dissimilarity induces larger paths between any pair of genes, on average, but with larger graph densities. Returning to the intensity of stress, the Spearman coefficient between $CPL$ and intensity of heat-shock stress is 0.95 ($Cosine$) and -0.74 ($GP$). Although these values do not show p-values below the 0.05 threshold, they are consistent with those reported in table~\ref{tab:Spearman correlation all measures} for the complete set of stresses.

Regarding the connection between the number of ESR genes in the top 1000 ranked genes according to the differential expression analysis (see Fig.~\ref{fig:Fig_intersect_top1000_inFS}) and the average path length, the rank correlation coefficient is $0.5$ and $-0.11$ for cosine and  GP dissimilarity, respectively. Therefore, we must conclude that stresses with a higher percentage of ESR genes with a larger differential expression response show more distinct expression patterns, if a cosine distance is used. 

Below we will resort to diffusion based properties summarizing both heterogeneity and connectivity represented by RMSTs. The amount of information preserved from the full weighted matrices will also be addressed by testing the correlation between node related entropy before and after sparsification with the RMST algorithm.

\subsection{Diffusion-based properties of stress-induced co-expression networks}\label{sec:Diff_Properties}

Our ultimate practical aim in this work is to use the full capacity of the RMST algorithm to provide the best substrate for finding clustering solutions of the expression patterns, as well as understanding cross-network organization features. Due to the nature of the community detection algorithm used in our work, and in order to perform a consistent analysis throughout the paper, we also evaluated the representation capacity of each RMST via the flow patterns on the network, thus augmenting the analysis performed above. Moreover, the diffusion properties across each network also help to distinguish each stress induced expression pattern, similarly to the heterogeneity property reported above. Below, we complement the analysis performed in the main text on diffusion-based properties.

Although the RMSTs do not share a substantial amount of links (see Fig.~\ref{fig:Fig_Hclust_Jaccard}), the full matrix of weights calculated for each stress shows considerable resemblances. In Fig.~\ref{fig:Correlation_WeiMat_allstresses} (A) and (B), we can observe that the Pearson correlation coefficients between the entries of each of the stress-induced weights matrices reaches values as high as 0.5. Across distance functions the Spearman correlation coefficient also shows considerably high values (see Fig.~\ref{fig:Correlation_WeiMat_allstresses} (C)). This informs us that the distinguishing features of the geometry of the data captured by the RMST algorithm allows extra crucial information that would not be immediately available, or would be difficult to extricate, had we relied on the full expression matrices for finding similar features or distinguishing aspects inherent to each stress response. 

If, on the other hand, we calculate the entropy rate and non-equilibrium entropy Spearman correlation coefficient across stresses and distance functions, we verify that the coefficients are much lower and less significant overall (see Fig.~\ref{fig:Fig_SpearmanCorr_allexp} (A), (C) and (E)). The cosine dissimilarity shows more significant values than $GP$, but does not allow higher $\rho_{S}$. See also Fig.~\ref{fig:Fig_SpearmanCorr_allexp_boxplot} (A) summarizing the results of Fig.~\ref{fig:Fig_SpearmanCorr_allexp}. Still, the non-equilibrium entropy allows for higher coefficients to be obtained (see~Fig.~\ref{fig:Fig_SpearmanCorr_allexp_boxplot}).

\subsubsection{Common stress response from entropy ranking cross-stress correlation}\label{sec:DiffusionRanksESR}

One interesting aspect of the results that should be noted relates to the set of $ESR$ genes, which were identified in the original study as a common response among stresses \cite{Gasch2000A}. If we perform the same correlation analysis as before, with entropy rate ranks as well as non-equilibrium entropy, the set of $\rho_{S}'s$ obtained points to interesting features common to every stress: there is a shift towards positive $\rho_{S}'s$ for $ESR$ genes for both distance functions when entropy rate ranks are computed, although not significant for $GP$ (Fig.~\ref{fig:Fig_SpearmanCorr_allexp_boxplot} (B)); for non-equilibrium entropy a wider range of rank correlation coefficients are generated but significance towards higher values is only observed for the $GP$ dissimilarity (Fig.~\ref{fig:Fig_SpearmanCorr_allexp_boxplot} (C)). Although the overall heterogeneity is correlated with the number of ESR genes in the top differentially expressed genes, and in addition the $\rho_{S}(\frac{h}{h_{max}},ESR)=-0.37$ ($Cosine$) and $-0.44$ ($GP$), the divergence in the response affects all genes. This leads to ESR genes having similar relative local heterogeneity profiles across stresses. It also points to nuanced ESR responses that might underlie up to a degree interesting aspects in cross-stress protection linked to the divergence in high-dimensional expression paths \cite{Gasch2000A,Gasch2007A,Berry2008A,Berry2011A}.

% ------------------------------------------------------------------------------------------------------------------------------------------
% ------------------------------------------------------------------------------------------------------------------------------------------
\section{Performance metrics for clustering methods}\label{sec:Clustering performance functions}

In order to quantify the average distance between genes in the clusters identified by the methods tested, we resorted to two sets of functions. The first set, composed of average intra-cluster homogeneity, inter-cluster separation and leave-one-out cross-validation, aims at understanding how close the shape of signals of genes clustered together really are. In addition, we also relied on another set of performance functions highlighting semantic similarity between gene ontology terms characterizing genes grouped together.

% ------------------------------------------------------------------------------------------------------------------------------------------
% ------------------------------------------------------------------------------------------------------------------------------------------
\subsection{Average intra-cluster homogeneity and inter-cluster separation}

The homogeneity of a partition reveals how close on average 2 genes clustered together are (see Eq.~\ref{eq:Eq_Hav}). Here we use the cosine similarity in order to measure how two time-course signals match. On the other hand the separation of a partition shows us how different on average pairs of genes from different clusters are (see Eq.~\ref{eq:Eq_Sav}). A good partition will have both high homogeneity and a very negative separation score.

\begin{eqnarray}
H_{av}=\frac{1}{K}\sum^{K}_{k=1}\frac{2}{n_{k}(n_{k}-1)}\sum_{\mathbf{g}_{i}(\mathbf{t}),\mathbf{g}_{j}(\mathbf{t})\in C_{k}}\frac{1+cos(\mathbf{g}_{i}(\mathbf{t}),\mathbf{g}_{j}(\mathbf{t}))}{2}\label{eq:Eq_Hav}
\end{eqnarray}

\begin{eqnarray}
S_{av}=\frac{1}{K}\sum^{K}_{k=1}\frac{2}{n_{k}(n-n_{k})}\sum_{\mathbf{g}_{i}(\mathbf{t}) \in C_{k}, \mathbf{g}_{j}(\mathbf{t}) \notin C_{k}}\frac{1+cos(\mathbf{g}_{i}(\mathbf{t}),\mathbf{g}_{j}(\mathbf{t}))}{2}\label{eq:Eq_Sav}
\end{eqnarray}

In the previous expression K stands for the number of clusters found for a specific partition, at a particular Markov time, and $C_{k}$ represents the clusters of a particular partition. As in previous sections, $\mathbf{t}=(t_{1},...,t_{N})$ and $\mathbf{g}_{i}(\mathbf{t})=(g_{1},...,g_{t_{N}})$ represents the expression profile for gene i.

% ------------------------------------------------------------------------------------------------------------------------------------------% ------------------------------------------------------------------------------------------------------------------------------------------
\subsection{Cross-validation of clusters}\label{sec:Cross-validation of clusters}

In order to quantify the quality of the clusters generated by the stability method at each Markov time, we also resorted once again to Gaussian process regression analysis and calculated the predictive log probability for each data point in each gene profile belonging to the cluster in question, a procedure known as Leave-one-out cross-validation ($LOOCV$). Although in certain cases this cross-validation technique analysis is computationally prohibitive, the fact that we use Gaussian process regression considerably simplifies the problem. The intra-cluster $LOOCV$ predictive probability is calculated according to Eq.~\ref{eq:Eq_LOOCV}.

\begin{eqnarray} 
 L_{LOOCV}({\bf t,g,\theta})=\sum^{N_{c}}_{i=1}\sum^{t_{N}}_{j=1}log\ \ p(g_{i,j}|{\bf t,g_{i,-j},g_{-i,j},\theta}) \label{eq:Eq_LOOCV}
\end{eqnarray}

In Eq.~\ref{eq:Eq_LOOCV} $log\ \ p(g_{i,j}|{\bf t,g_{i,-j},g_{-i,j},\theta})$ is determined according  to Eq.~\ref{eq:Eq_predprob}, $\mathbf{t}=(t_{1},...,t_{N})$, and $\mathbf{g_{i}}=(g_{1},...,g_{j},...,g_{t_{N}})$ represents the time-series for gene i, collected at $t_{N}$ time-points. Therefore $L_{LOOCV}$ is calculated by successfully evaluating the predictive probability of expression level $g_{i,j}$, i.e. the data collected for gene i at instant j, given all the data in a specific cluster at all instants for all other genes $\mathbf{g_{-i,j}}$ and the rest of the data for gene i collected at any other time-points $\mathbf{g_{i,-j}}$. This calculation is performed for each gene in a cluster, c, made of $N_{c}$ genes.

\begin{eqnarray} 
 log\ \ p(g_{i,j}|{\bf t,g_{i,-j},g_{-i,j},\theta})=-\frac{1}{2} ln \sigma_{i,j}^{2}-\frac{1}{2} \frac{(g_{i,j}-\mu_{i,j})^{2}}{\sigma_{i,j}^{2}}-\frac{1}{2} ln 2\pi \label{eq:Eq_predprob}
\end{eqnarray}

 Both $\sigma_{i,j}$ and $\mu_{i,j}$ can be calculated through the standard prediction formulas for Gaussian process regression \cite{Rasmussen2006A} with the square-exponential kernel {\bf $K_{t}$} (see Eq.~\ref{eq:covariance_kernel}). Subsequently, in order to quantify the quality of a partition at a specific Markov time we calculate the average $LOOCV_{av}=\frac{1}{K}\sum^{K}_{k=1}L_{LOOCV}$ predictive probability over the clusters characterizing the partition in question.

% ------------------------------------------------------------------------------------------------------------------------------------------
% ------------------------------------------------------------------------------------------------------------------------------------------
\subsection{Average intra-cluster gene semantic similarity}

Gene ontology term overlap (GOTO) \cite{Mistry2008A} similarity score is a measure of the relative number of GO terms shared between pairs of genes clustered together (see Eq.~\ref{eq:Eq_GOTO}).

\begin{eqnarray}
\overline{GOTO_{k}}=\frac{2}{n_{k}(n_{k}-1)}\sum_{g_{i},g_{j}\in C_{k}}\left|annot_{i} \cap annot_{j}\right|\label{eq:Eq_GOTO}
\end{eqnarray}

In Eq.~\ref{eq:Eq_GOTO} $n_{k}$ is the number of genes in cluster or community $C_{k}$, and $\left|annot_{i} \cap annot_{j}\right|$ stands for the number of GO terms that a pair of genes in cluster $k$ have in common. The annotation of each gene, $annot_{i,j}$, includes all the terms from the leaves of the hierarchy up to, yet excluding its root \cite{Mistry2008A}. In this way if two genes are involved in the same biological processes or have similar molecular functions, the score reflects this common ground up to higher levels of biological information. In this work we calculated the scores for biological process (bp), molecular function (mf) and cellular compartment (cc) annotations. An overall measure of the clustering quality in terms of semantic similarity can be determined by calculating the weighted average of the $\overline{GOTO_{k}}$ scores per cluster/community (see Eq.~\ref{eq:Eq_GOTOav}) (see also \cite{Kirk2012A}).

\begin{eqnarray}
GOTO_{av}=\sum_{k}\left(\frac{n_{k}}{n}\right)\overline{GOTO_{k}}\label{eq:Eq_GOTOav}
\end{eqnarray}

Unlike other semantic similarity measures for cluster validation, e.g. the biological homogeneity index \cite{Datta2006A,Brock2008A}, the $GOTO_{av}$ score shows more specificity and allows for a more informative measure due to the fact that if two genes share more specific GO terms then they will score higher. In addition, the $GOTO_{av}$ score also allows for a quicker calculation of semantic similarity than, for instant, information content metrics \cite{Mistry2008A,Lord2003A}.

% ------------------------------------------------------------------------------------------------------------------------------------------
% ------------------------------------------------------------------------------------------------------------------------------------------
\section{Performance of clustering analysis with directed co-expression networks:further results}\label{sec:PerformanceRMSTplusCSI}

The performance associated with the set of simulations when the weights matrix in the causal co-expression network are those computed through Eq.~8 (Main Text) is reported in table~\ref{tab:PerformanceRMSTplusCSI}. Overall, the method improves considerably across most measures for the RMSTs originally determined from a $GP$ dissimilarity function.

\begin{table}[ht]
\begin{center}
\begin{tabular}{c|c|c}
 
Function        &   Cosine (dir.) - Cosine (dir.CSI)    & GP (dir.) - GP (dir.CSI)\\
\hline 
$H_{av}$        &   {\bf 0.76} - 0.24 &  0.38 - {\bf 0.62}  \\

$S_{av}$        &   0.03 - {\bf 0.96} &  0.005 - {\bf 0.95} \\

$LOOCV_{av}$    &   0.46 - {\bf 0.54} &  {\bf 0.58} - 0.42  \\

$GOTObp_{av}$   &   {\bf 0.59} - 0.41 &  0.42 - {\bf 0.58}  \\

$GOTOcc_{av}$   &   {\bf 0.55} - 0.44 &  0.41 - {\bf 0.59}  \\

$GOTOmf_{av}$   &   {\bf 0.58} - 0.41 &  0.39 - {\bf 0.61}  \\

\end{tabular}
\caption[Performance of Markov stability clustering analysis, weights calculated according to the GP distance (dir.) vs weights calculated according to Eq.~8 (dir.CSI)]{\label{tab:PerformanceRMSTplusCSI} Performance of Markov stability clustering analysis, weights calculated according to the GP distance (dir.) vs weights calculated according to Eq.~8 (dir.CSI). The results presented here are the proportions of clustering solutions where a specific distance function, of each pair, induces better results according to the selected performance measures. Cases that do not add up to 1 exclude instances where performance was equal.}
\end{center}
\end{table}

% ------------------------------------------------------------------------------------------------------------------------------------------
% ------------------------------------------------------------------------------------------------------------------------------------------

\section{Closest partition to the Bayesian Hierarchical Clustering best solution}\label{sec:CloseToBestBHC}

The BHC algorithm determines the best partition for each stress with a considerable number of clusters, also over partitioning the data sets. In order to understand at which resolution we should observe the partitions determined by Markov stability analysis that would be the closest to those proposed by the BHC, we calculated 2 information based distance functions, the normalized variation of information ($NVI$) and the adjusted mutual information ($AMI$, see Eq.~6) \cite{Vinh2009An,Vinh2010A}, for each stress, between the clustering solutions proposed by both methods (see Fig.~\ref{fig:DistancetoBHC_allstresses_allstresses}). 

The normalized variation of information despite being a metric is sensitive to the number of communities \cite{Vinh2009An,Vinh2010A}. The AMI function, on the other hand, does not show the same dependence, and identifies with greater accuracy partitions determined by Markov stability with similar community organization as those found by the BHC method (Fig.~\ref{fig:DistancetoBHC_allstresses_allstresses} (D)). Overall, we would have to look at very high resolutions (low Markov times) in order to reach a partition solution similar to those proposed by the BHC algorithm.

\section{Multi-scale organization across stresses}\label{sec:Cross-StressStruct}

The set of networks reported in the section of the main text analysing the cross-stress similarities between partition solution found for each stress are shown below:

\begin{itemize}

\item The network highlighting cross-stress features for nodes representing clustering solutions with a number of groups above 2 and below 500 can be seen in Fig.~\ref{fig:Fig_Part_RMST_N}; 

\item Details of stability function and robustness associated with each node in Fig.~6 are coloured accordingly in the spatial embeddings of Fig.~\ref{fig:Fig_Part_RMST_Nupto50_VI_S};

\item Details of the spatial embedding shown in Fig.~6 (B) can be explored in Fig.~\ref{fig:Fig_Part_RMST_Nupto50_ZOOM};

\item The joint partition described in the main text is shown in Fig.~\ref{fig:Fig_Part_proj_Nupto50_VIbest}.
\end{itemize}

% ------------------------------------------------------------------------------------------------------------------------------------------
% ------------------------------------------------------------------------------------------------------------------------------------------

\section{Including other sources of information: the case of physical and genetic interaction networks}\label{sec:BIOGRID}

Given that we are comparing partition solutions via information based similarity functions, the same procedure can be applied to any other sources of information about the same set of genes used in the analysis presented above. We chose to enhance the multi-resolution analysis presented before with information collected from the $BioGRID$ database ($http://thebiogrid.org$). The number of unique genes in the $BioGRID$ {\it S.cerevisiae} network totals 6504 with 225,753 non-redundant edges \footnote{Effectively, in the information collected from $BioGRID$ (non-redundant edges), each unique combination of interactors are counted as a single interaction, regardless of directionality, the experimental system used for determining the edge and publication.}, of which 81,115 are physical interactions and the remaining represent genetic interactions.

Physical interactions on $BioGRID$ are interpreted as arising in essays determining local aspects of interaction, from affinity capture essays to those evaluating biochemical activity, e.g. phosphorylation, or protein-protein interactions \footnote{See $http://wiki.thebiogrid.org/doku.php/experimental_systems$ for details of experimental systems}. Genetic interactions, on the other hand, represent a more high-level representation of relationships between genes, identified by mutation studies that clarify the connection between genes and pathways not at a local level but at a functional level associated with probability of survival \cite{Costanzo2010A}. Therefore, these two sources of additional information on each gene may provide a clarifying perspective, from a network structure angle, of the underlying mechanisms at play under each stress. A finer analysis could be performed where, for instance, the physical interaction subset would be further reduced by focussing exclusively on phosphorylation or protein-protein interactions. 

There are other datasets that provide further information linked more directly to co-expression networks, e.g. $YEASTRACT$ ($http://www.yeastract.com$), where a regulatory matrix has been collated with both activating and inhibitory interactions listed. Despite this being an excellent source of information, the connected largest core of genes common to this dataset and the differentially expressed genes used in the analysis presented above was comparatively small. Further analysis will be performed elsewhere.

\subsection{BIOGRID physical interaction}

By determining the multi-scale structure of the largest connected core of 5544 genes belonging to the physical interaction network ($PIN$), we are capable of once more finding cross-dataset partition similarities by resorting to the similarity function $AMI$. The networks used in this section are all undirected.

Throughout the work presented until now the number of genes used for stress induced multi-scale analysis was 2000. Therefore, for consistency purposes, we intersected the largest connected core of genes from the $BioGRID$ PIN with the set of 2000 differentially expressed genes from the Gash and coworkers microarray dataset, resulting in a set of 1659 genes. We must emphasize, nevertheless, that the multi-scale organization of the $BioGRID$ PIN was found with the whole set of genes belonging to the largest core, as stated above. Effectively, narrowing down the attention to the common 1659 genes represented in both datasets amounts to extracting from the $BioGRID$ multi-scale partition solution membership or indicator matrix, only the rows corresponding to genes common to the differentially expressed set.

Due to the discrepancy in the order of magnitude between stress-stress and stress-PIN $AMI$ values, we decided to combine the RMST structure obtained before for co-expression networks (see Fig.~\ref{fig:Fig_Part_RMST_N}), with an MST stemming from applying the $AMI$ function to the augmented similarity matrix. This avoids problems related to overdense networks being generated if the criterion underlying the RMST algorithm is applied to matrices where a drastic drop of several orders of $AMI$ magnitude is present. 

In Fig.~\ref{fig:Fig_Links_BIOGRIDNet}, we can verify that the partition solutions of the physical interaction $BioGRID$ network have selective connectivity to the rest of the partition space RMST. For example, focusing on the network in partition space arising from selecting solutions with no more than 50 clusters we verify that only $Diam.$, $HP$, $HS25-37$ and $HS37-25$, have links between partition solutions representing time-expression patterns and partition solutions of the $BioGRID$ PIN. Amongst the stresses just mentioned, that with the highest number of connections and average weight, is once again diamide. This indicates that diamide induces an expression pattern that to an extent is determined by the local interactions of proteins, via complex formation or other type of reaction represented as physical interaction in $BioGRID$ dataset.

The identification of local structures as well as local heterogeneity of connecting weights has been an important tool of analysis of protein-protein networks combined with expression correlation matrices. One of the most cited frameworks is that of signalling entropy \cite{Teschendorff2014A}, which allows for identification of both local hotspots and global properties differentiating samples collected in different environments. Effectively, the combination of PIN networks with expression patterns, by weighting the PIN links with a measure expression proximity, assumes an implicit mapping between dynamics and physical structure of intra-cellular networks. Contrary to this approach, we evaluate the proximity between different datasets by taking into account its multi-scale organization. Consequently, if there is a link between clustering solutions in different datasets or data types, i.e. $BioGRID$ PIN and any other co-expression network solution, this must indicate a common dynamical pattern emerging from a local physical interaction pattern under a particular stress. 

The same analysis could have been generated with the partitions found for each stress when causality is imposed on each edge of the expression space RMSTs. Yet, here the objective is to show only general features of the partition space of combined networks. Details about the clusters obtained for the $BioGRID$ PIN as well as the $BioGRID$ genetic interaction networks (GIN) will be explored elsewhere.

\subsection{BIOGRID genetic interaction}

Following the same steps as those highlighted above for the $BioGRID$ PIN network, we also analysed the genetic interactions largest connected core structure network (5327 genes) part of the $BioGRIG$ database for {\it S.cerevisiae}. As was observed before, the stress diamide has the highest number of connections with the set of partition solutions in expression space (see Fig.~\ref{fig:Fig_Links_BIOGRIDNet} (B)). Contrary to the results observed for the $BioGRID$ PIN (see Fig.~\ref{fig:Fig_Links_BIOGRIDNet} (A)), the only stress that does not connect at all with solutions for the genetic interaction network is $HS29-33(I)$. Nevertheless, this stress belongs to the same category of stresses, at least partially, of that not connected to any of the solutions of the $BioGRID$ PIN, i.e. $HS29-33 (II)$ (see Fig.~\ref{fig:Fig_Links_BIOGRIDNet} (A)). The former is a pure mild temperature stress and the latter combines the former with osmotic stress.

Overall, the Spearman correlation between the ranks observed for the edge properties recorded in Fig.~\ref{fig:Fig_Links_BIOGRIDNet} in both $BioGRID$ subsets, PIN and GIN, is $-0.1455$ and $ 0.3659$, for number of edges and average edge weight, respectively, although not significant at a $0.05$ level.  Nevertheless, an interesting aspect is observed for diamide, hydrogen peroxide, high amplitude temperature shock ($HS25-37$) and hyper-osmotic stress, which correspond to intense stress conditions: they are in the top ranks, i.e. under these stresses, according to the analysis presented here, the expression dynamics arise closely from both local physical interaction communities as well as functional relationships captured by genetic interactions experiments \cite{Costanzo2010A}.

% ------------------------------------------------------------------------------------------------------------------------------------------
% ------------------------------------------------------------------------------------------------------------------------------------------

\section{Overlap between sets of genes differentially expressed under different stresses} \label{sec:Overlap_diffexpr_genes}

In order to find the clusters characterizing the response to each of the stresses we selected only genes that had a high score according to the test for differential expression outlined above. The intersection between differentially expressed genes per stress is different between pairs of stresses (see Fig.~\ref{fig:Fig_intersect_top1000}). A set of Venn diagrams (Figs.~\ref{fig:Fig_intersect_top1000HS_Venn}, \ref{fig:Fig_intersect_top1000HSOS_Venn} and \ref{fig:Fig_intersect_top1000_Venn}) shows the intersection between the top 1000 genes scored according to the one-sample test presented above. For heat shock we observe that only one gene is common to all temperature related stresses. The largest intersection between pairs of temperature related stresses corresponds to $HS25-37$ and $HS30-37$, thus indicating amplitude related aspects associated with stress. In Fig.~\ref{fig:Fig_intersect_top1000HSOS_Venn} we also show the intersection between the mild temperature shock stresses and the osmotic stress. Once again, there is only one gene associated with the intersection of all stresses. Surprisingly, the largest intersection between the mild heat shock stresses and the hyper-osmotic stress happens with $HS29-33(I)$. We would expect that the largest set of common differentially expressed genes would occur between the hyper-osmotic stress and $HS29-33(II)$. The latter can be seen as combinatorial stress, where cells where grown in a higher osmolarity medium and subsequently exposed to mild heat shock \cite{Gasch2000A}. This surprising result might highlight different stages of stress response where, despite the same stresses being present, there is an early and late response characteristic of each stress \cite{Gasch2000A}. Also interesting is the absence of a striking intersection between the list of genes elicited by each stress in the 3 stress set highlighted here. Interestingly, from the 172 genes common to $Hyper-osmotic$ and $HS29-33(I)$, only 27 belong to the 1000 gene group previously referred to as the Environmental-Stress-Response (ESR) \cite{Gasch2000A,Gasch2002A,Gasch2007A}. In fact, this number is higher than that recorded for the intersection between $Hyper-osmotic$ and $HS29-33(II)$. Common to all 3 stresses there are only 26 genes, of which only 1 belong to the ESR list (see Figs.~\ref{fig:Fig_intersect_top1000} and~\ref{fig:Fig_intersect_top1000ESR_Venn}). A noticeable feature of the 1000 gene set, per stress, selected by the one-sample test presented in the differential analysis section, is the absence of a common overlap between all sets.

Overall, as was mentioned above, we selected initially the union of the top 1000 genes per stress and subsequently retained only the 2000 that had the highest variance across stresses. Here we should further add that the whole set of genes was normalized with respect to controls and zero transformed, i.e. divided by the the expression at instant t=0. Therefore, any analysis is done on the changes with respect to the initial state,  which makes the variance selection criterion a valid one. The largest contributor to the final set of genes for analysis is the stress diamide with 648 genes from the original top 1000 being present in the final selection (see Fig.~\ref{fig:Fig_intersect_top1000_inFS}). 

This filtering process reduces significantly the amount of genes to be clustered and allows for comparison of the responses of the same genes across stresses. Also, the number of genes selected allows for a considerable set of responses to be retained including those responsible for the environmental stress response \cite{Gasch2000A}. 

All Venn diagrams presented were generated with the VennDiagram R package ($http://cran.r-project.org/web/packages/VennDiagram/index.html$).

%--------------------------------------------------------------------------------------------------------------------------------------------
%--------------------------------------------------------------------------------------------------------------------------------------------
%                                          References

%--------------------------------------------------------------------------------------------------------------------------------------------
%--------------------------------------------------------------------------------------------------------------------------------------------
\bibliographystyle{unsrt}

\newpage

%--------------------------------------------------------------------------------------------------------------------------------------------
%--------------------------------------------------------------------------------------------------------------------------------------------
%                                          All Supplementary Figures

%--------------------------------------------------------------------------------------------------------------------------------------------
%--------------------------------------------------------------------------------------------------------------------------------------------

\begin{figure}[ht]
\caption[Hierarchical clustering of RMST co-expression networks generated with a $Cosine$ (A) and $GP$ distance function (B)]{\label{fig:Fig_Hclust_Jaccard} Hierarchical clustering of RMST co-expression networks generated with a $Cosine$ (A) and $GP$ distance function (B). For both, the distance between networks was determined via a simple Jaccard distance, i.e. $ 1-\frac{|E_{i} \cap E_{j}|}{|E_{i} \cup E_{j}|}$, where $E_{i}$ stands for adjacency matrix of RMST co-expression network i and $\frac{|E_{i} \cap E_{j}|}{|E_{i} \cup E_{j}|}$ for the proportion of the number of edges shared between networks i and j with respect to all the edges present in both.}
\end{figure}

%--------------------------------------------------------------------------------------------------------------------------------------------
%--------------------------------------------------------------------------------------------------------------------------------------------

%--------------------------------------------------------------------------------------------------------------------------------------------
%--------------------------------------------------------------------------------------------------------------------------------------------

\begin{figure}[ht]
\caption[Co-expression network density and characteristic path length for different distance functions]{\label{fig:Fig_RMSTsDensity} Co-expression network density and characteristic path length for different distance functions. (A) Graph density per stress and (B) Characteristic path length for the cosine and Gaussian process based distance function.}
\end{figure}

%--------------------------------------------------------------------------------------------------------------------------------------------
%--------------------------------------------------------------------------------------------------------------------------------------------

\begin{figure}[ht]
\caption[Co-expression networks degree for different distance functions]{\label{fig:Fig_RMSTsDegree} Co-expression networks degree for different distance functions. Degree distribution per stress for cosine and Gaussian process based distance function.}
\end{figure}

%--------------------------------------------------------------------------------------------------------------------------------------------
%--------------------------------------------------------------------------------------------------------------------------------------------

\begin{figure}[ht]
\caption[Coefficient of variation for weights matrices per stress]{\label{fig:Heterogeneity_allstresses_CV} Coefficient of variation for weights matrices per stress. (A) Full matrices. (B) RMST weighted matrices.}
\end{figure}

%--------------------------------------------------------------------------------------------------------------------------------------------
%--------------------------------------------------------------------------------------------------------------------------------------------

\begin{figure}[ht]
\caption[Correlation between weights matrices across stresses]{\label{fig:Correlation_WeiMat_allstresses} Correlation between weights matrices across stresses. (A) Cross-stress Pearson correlation for $Cosine$ and (B) $GP$ distance function. (C) Cross-stress, cross-distance Spearman correlation function. (D) Diagonal correlation values in (C).}
\end{figure}

%--------------------------------------------------------------------------------------------------------------------------------------------
%--------------------------------------------------------------------------------------------------------------------------------------------

\begin{figure}[ht]
\caption[Entropy rate for each stress-induced RMST]{\label{fig:Fig_EntropyRate_Suppl} Entropy rate for each stress-induced RMST. (A) Entropy rate normalized by the optimal entropy rate corresponding to each weighted structure. (B) Normalized entropy rate for each weighted full matrix (normalization factor is the entropy rate of a full matrix with equal weights). (C) Rank correlation between entropy rate values per node before and after applying the RMST algorithm. (D) P-values corresponding to Spearman correlation coefficients shown in (C).}
\end{figure}

%--------------------------------------------------------------------------------------------------------------------------------------------
%--------------------------------------------------------------------------------------------------------------------------------------------

\begin{figure}[ht]
\centering
\caption[Spearman correlation across stresses and distance functions for entropy rates per node]{\label{fig:Fig_SpearmanCorr_allexp} Spearman correlation across stresses and distance functions for entropy rates per node. (A) $Cosine$ distance function. (B) P-values for values in (A). (C) $GP$ distance function. (D) P-values for values in (C). (E) Cross- distance correlation. (F) P-values for values in (E).}
\end{figure}

%--------------------------------------------------------------------------------------------------------------------------------------------
%--------------------------------------------------------------------------------------------------------------------------------------------

\begin{figure}[ht]
\caption[Boxplot of Spearman correlations across stresses and distances functions for entropy rates and non-equilibrium entropy]{\label{fig:Fig_SpearmanCorr_allexp_boxplot} Box plot of Spearman correlations across stresses and distances functions for entropy rates and non-equilibrium entropy. (A) Summary of Spearman correlation matrices shown in Fig.~\ref{fig:Fig_SpearmanCorr_allexp}. (B) Distinction, with respect to entropy rate, between Environmental Stress Response (ESR) genes and others. Wilcoxon signed rank statistic reveals that the distinction is significant for $Cosine$ dissimilarity ($p-value=3.3714e-06 << 0.05$) but not for GP. (C) Distinction, with respect to non-equilibrium entropy, between ESR genes and others. Wilcoxon signed rank statistic shows significance for GP dissimilarity ($p-value=0.0274 < 0.05$) but not for $Cosine$.}
\end{figure}

%--------------------------------------------------------------------------------------------------------------------------------------------
%--------------------------------------------------------------------------------------------------------------------------------------------

\begin{figure}[ht]
\caption[Number of clusters in the best Bayesian Hierarchical Clustering solutions for all stresses and closest Markov stability partition determined by information theory based distance functions]{\label{fig:DistancetoBHC_allstresses_allstresses} Number of clusters in the best Bayesian Hierarchical Clustering solutions for all stresses and closest Markov stability partition determined by information theory based distance functions. (A) Number of clusters in best BHC partition, compared to the closest Markov stability partition determined by variation of information with respect to the partitions across all Markov times (see also (B)). (B) Variation of information between the best BHC partition for stress $HS25-37$ and the solutions obtained via Markov stability. (C) Numbers of clusters in best BHC compared to the closest Markov stability partition determined by adjusted mutual information with respect to the partitions across all Markov times (see also (D)). (D) Adjusted mutual information between the best BHC partition for stress HS25-37 and the solutions obtained via Markov stability.}
\end{figure}

%--------------------------------------------------------------------------------------------------------------------------------------------
%--------------------------------------------------------------------------------------------------------------------------------------------

\begin{figure}[ht]
\caption[Network representation of cross stress-induced co-expression network multi-scale community structure, with emphasis on cluster number, for solutions found by using $GP$ dissimilarity in expression space]{\label{fig:Fig_Part_RMST_N} Network representation of cross stress-induced co-expression network multi-scale community structure, with emphasis on cluster number, for solutions found by using a $GP$ dissimilarity function in expression space. The function used for calculating the distance between partitions was the adjusted mutual information ($AMI$). (A) Nodes coloured by the number of communities found by Markov stability analysis in expression space, intensity ascending from blue to red. Spatial embedding performed with the Fruchterman-Reingold method available in Gephi. Nodes represent clustering solutions with a maximum of 500 clusters and a minimum 2. (B) Details of edges connecting nodes in (A) for solutions with at most 50 clusters. Colour indicates weights, i.e. $AMI$ intensity ascending from blue to red.}
\end{figure}

%--------------------------------------------------------------------------------------------------------------------------------------------
%--------------------------------------------------------------------------------------------------------------------------------------------

\begin{figure}[ht]
\caption[Detail of RMST of partitions with up to 50 clusters (see also Fig.~6 (B) in main text). Network with nodes coloured, from blue to red, by number of clusters (also represented) found in expression space. Node size is proportional to degree]{\label{fig:Fig_Part_RMST_Nupto50_ZOOM} Detail of RMST of partitions with up to 50 clusters (see also Fig.~6 (B) in main text). Network with nodes coloured, from blue to red, by number of clusters (also represented) found in expression space. Node size is proportional to degree. Edge thickness is proportional to weight, in this case the $AMI$. Spatial embedding with force based algorithm available in Gephi.}
\end{figure}

%--------------------------------------------------------------------------------------------------------------------------------------------
%--------------------------------------------------------------------------------------------------------------------------------------------

\begin{figure}[ht]
\caption[Detail of network of partitions with up to 50 clusters, stability value and robustness]{\label{fig:Fig_Part_RMST_Nupto50_VI_S} Detail of network of partitions with up to 50 clusters, stability value and robustness. (A) Stability. (B) Robustness of solution, i.e. average VI between 100 Louvain solutions. Numbers correspond to numbers of clusters (see also Fig.~6 in main text). Node size is proportional to degree. Edge thickness is proportional to weight, in this case the $AMI$. Spatial embedding with force based algorithm in Gephi.}
\end{figure}
%--------------------------------------------------------------------------------------------------------------------------------------------
%--------------------------------------------------------------------------------------------------------------------------------------------

\begin{figure}[ht]
\caption[Joint partition selection]{\label{fig:Fig_Part_proj_Nupto50_VIbest} Joint partition selection. (A) Joint partition selected from the graph represented in Fig.~6 (main text). (A) By retaining only nodes and respective connections that were found to be the most robust solution, and (B) Node colour representing number of clusters for the respective node/partition in expression space. Node size represents degree in original network (see Fig.~6). $\gamma=0.0022$ (see Eq.~6 in main text).}
\end{figure}

%--------------------------------------------------------------------------------------------------------------------------------------------
%--------------------------------------------------------------------------------------------------------------------------------------------

\begin{figure}[ht]
\caption[Comparison between different sources of information for the same gene]{\label{fig:Fig_Links_BIOGRIDNet} Comparison between different sources of information for the same gene. (A) Edges between the $BioGRID$ physical interaction largest core network partitions and stress-induced co-expression network partitions. (B) Edges between the $BioGRID$ genetic interaction largest core network partitions and stress-induced co-expression network partitions. Colour-scale represents $AMI$. $N$ stands for the number of communities registered for each network partition in each vertex connected by the edges. $GP$ expression space distance function.}
\end{figure}

%--------------------------------------------------------------------------------------------------------------------------------------------
%--------------------------------------------------------------------------------------------------------------------------------------------

\begin{figure}[ht]
\caption[Edge properties between $BioGRID$ physical and genetic interaction networks and stress-induced co-expression networks]{\label{fig:Fig_Part_InterStressBIOGRID_edges} Edge properties between $BioGRID$ physical and genetic interaction networks and stress-induced co-expression networks. (A) Number of edges. (B) Average edge weight strength.}
\end{figure}

%--------------------------------------------------------------------------------------------------------------------------------------------
%--------------------------------------------------------------------------------------------------------------------------------------------

\begin{figure}[ht]
%\centering
%\includegraphics[width=0.6\textwidth]{FigS16.jpg}
\caption[Pairwise intersection between sets of top 1000 genes selected by one-sample gaussian process differential expression analysis]{\label{fig:Fig_intersect_top1000} Pairwise intersection between sets of top 1000 genes selected by one-sample Gaussian process differential expression analysis. (A) All 1000 genes in each set. (B) Focus on ESR genes.}
\end{figure}

%--------------------------------------------------------------------------------------------------------------------------------------------
%--------------------------------------------------------------------------------------------------------------------------------------------

\begin{figure}[ht]
%\centering
%\includegraphics[width=0.6\textwidth]{FigS17.jpg}
\caption[Number of genes belonging to each top 1000 gene list per stress in the final 2000 gene selected set]{\label{fig:Fig_intersect_top1000_inFS}. Number of genes belonging to each top 1000 gene list per stress in the final 2000 gene selected set. (A) All 1000 genes. (B) Focus on ESR genes.}
\end{figure}

%--------------------------------------------------------------------------------------------------------------------------------------------
%--------------------------------------------------------------------------------------------------------------------------------------------

\begin{figure}[ht]
%\centering
%\includegraphics[width=1\textwidth]{FigS18.jpg}
\caption[Venn diagram for intersection between top 1000 gene sets, per stress, selected by one-sample gaussian process differential expression analysis. Focus on temperature related stresses]{\label{fig:Fig_intersect_top1000HS_Venn} Venn diagram for intersection between top 1000 gene sets, per stress, selected by one-sample Gaussian process differential expression analysis. Focus on temperature related stresses.}
\end{figure}

%--------------------------------------------------------------------------------------------------------------------------------------------
%--------------------------------------------------------------------------------------------------------------------------------------------

\begin{figure}[ht]
%\centering
%\includegraphics[width=1\textwidth]{FigS19.jpg}
\caption[Venn diagram for intersection between top 1000 gene sets, per stress, selected by one-sample gaussian process differential expression analysis. Focus on mild temperature and osmotic related stresses]{\label{fig:Fig_intersect_top1000HSOS_Venn} Venn diagram for intersection between top 1000 gene sets, per stress, selected by one-sample Gaussian process differential expression analysis. Focus on mild temperature and osmotic related stresses.}
\end{figure}

%--------------------------------------------------------------------------------------------------------------------------------------------
%--------------------------------------------------------------------------------------------------------------------------------------------

\begin{figure}[ht]
%\centering
%\includegraphics[width=1\textwidth]{FigS20.jpg}
\caption[Venn diagram for intersection between top 1000 gene sets, per stress, selected by one-sample gaussian process differential expression analysis. Focus on temperature, osmotic and oxidative related stresses.]{\label{fig:Fig_intersect_top1000_Venn} Venn diagram for intersection between top 1000 gene sets, per stress, selected by one-sample Gaussian process differential expression analysis. Focus on temperature, osmotic and oxidative related stresses.}
\end{figure}

%--------------------------------------------------------------------------------------------------------------------------------------------
%--------------------------------------------------------------------------------------------------------------------------------------------

\begin{figure}[ht]
%\centering
%\includegraphics[width=0.7\textwidth]{FigS21.jpg}
\caption[Venn diagram for intersection between ESR genes in top 1000 gene sets, per stress, selected by one-sample Gaussian process differential expression analysis]{\label{fig:Fig_intersect_top1000ESR_Venn} Venn diagram for intersection between ESR genes in top 1000 gene sets, per stress, selected by one-sample Gaussian process differential expression analysis. (A) Focus on temperature and oxidative related stresses. (B) Focus on mild temperature and osmotic related stresses.}
\end{figure}

%--------------------------------------------------------------------------------------------------------------------------------------------
%--------------------------------------------------------------------------------------------------------------------------------------------

\end{document}